\documentclass[a4paper]{article}

\usepackage{amsfonts,amssymb,amsmath}
\usepackage{graphicx,graphics}
\usepackage{tabularx,hhline}
\usepackage{tikz}
\newcommand{\ie}{{ \textit i.e. }}

\renewcommand{\=}{\,=\,}
\newcommand{\+}{\,+\,}
\renewcommand{\-}{\,-\,}

\newcommand{\2}{2 \textit{in} - 2 \textit{out} }
\newcommand{\3}{3 \textit{in} - 1 \textit{out} }
\newcommand{\4}{3 \textit{out} - 1 \textit{in} }

\title{\textbf{Magnetic Monopole Dynamics in Spin Ice}}
\author{L. D. C. Jaubert$^{\ast}$  and P. C. W. Holdsworth$^{\dag}$}

\begin{document}

\maketitle

$^{\ast}$Max-Plack-Institut f\"ur Physik komplexer Systeme, 01187 Dresden, Germany.\\

$^{\dag}$Universit\'{e} de Lyon, Laboratoire de Physique, \'{E}cole Normale Sup\'{e}rieure de Lyon, 46 All\'{e}e d'Italie, 69364 Lyon cedex 07, France.\\

\date{\today}

\begin{abstract}

One of the most remarkable examples of emergent quasi-particles, is that of the "fractionalization" of magnetic dipoles in the low energy configurations of materials known as "spin ice", into free and unconfined magnetic monopoles interacting via Coulomb's $1/r$ law [Castelnovo {\it et. al.}, Nature, {\bf 451}, 42-45 (2008)].   Recent experiments have shown that a Coulomb gas of magnetic charges really does exist at low temperature in these materials 
and this discovery provides a new perspective on otherwise largely inaccessible phenomenology. In this paper, after a review of the different spin ice models, we present detailed results describing the diffusive dynamics of monopole particles starting both from the dipolar spin ice model and directly from a Coulomb gas within the grand canonical ensemble. The diffusive quasi-particle dynamics of real spin ice materials within "quantum tunneling" regime is modeled with Metropolis dynamics, with the particles constrained to move along an underlying network of oriented paths, which are classical analogues of the Dirac strings connecting pairs of Dirac monopoles.  

\end{abstract}

\section{Introduction}

\begin{figure} [ht]
\centering{\includegraphics[scale=0.5]{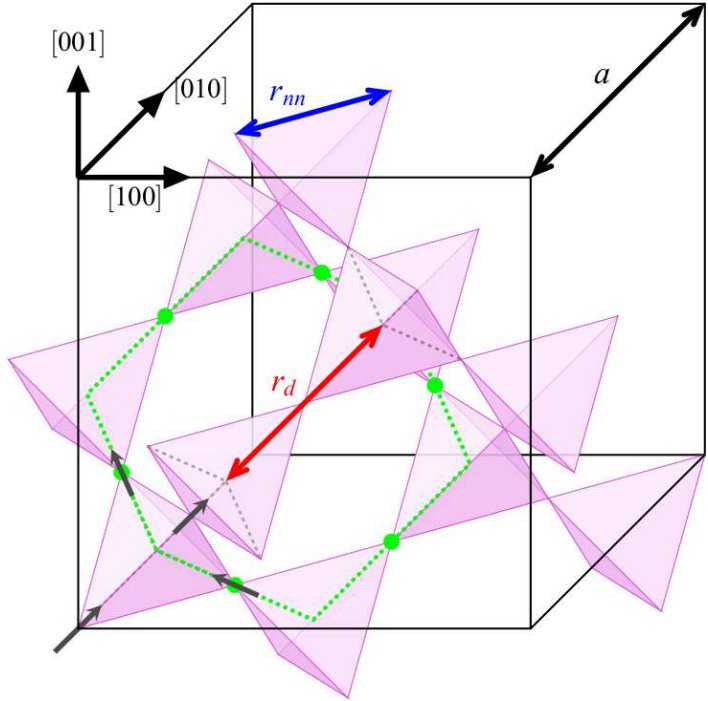}}
\caption[$\;\;$ Pyrochlore lattice]{\textbf{Pyrochlore lattice:} The spins are located on the corner of every tetrahedra and are fixed along their local [111] axis represented by the dashed lines. There are two types of tetrahedra that we shall call the \textit{down} tetrahedra (bottom left with the four spins) and the \textit{up} tetrahedra. Each \textit{down} tetrahedra is connected to four \textit{up} ones, and vice-versa. The cube represents a unit cell with 8 tetrahedra (four of each kind) and 16 spins, and defines the [100] ($x$), [010] ($y$) and [001] ($z$) axes. We introduce the length of the unit cell $a\approx10$ \AA$\,$ as well as the distance between nearest neighbour $r_{nn}=\frac{\sqrt{2}}{4}\,a\approx3.5$ \AA$\;$and between the centres of two connected tetrahedra $r_{d}=\frac{\sqrt{3}}{4}\,a\approx4.3$ \AA.  The smallest close loop encompasses 6 spins (see green dotted loop).}
\label{fig:intro:pyrochlore}
\end{figure}

Spin ice materials\cite{Harris97a,Gingras11a}  form part of a series of rare-earth oxide insulator R$_{2}$M$_{2}$O$_{7}$ with space group $Fd\overline{3}m$ where R$^{3+}$  is a magnetic  (Dy$^{3+}$ and Ho$^{3+}$) and M$^{4+}$ a non-magnetic (Ti$^{4+}$ or Sn$^{4+}$) ion. The cations sit on the vertices of two interpenetrating pyrochlore lattices formed by corner-sharing tetrahedra (see figure~\ref{fig:intro:pyrochlore}). The well-established members of this family are Dy$_{2}$Ti$_{2}$O$_{7}$ , Ho$_{2}$Ti$_{2}$O$_{7}$ and Ho$_{2}$Sn$_{2}$O$_{7}$, the latter being less studied as it is only available in polycrystalline form.
Other potential candidates are Dy$_{2}$Sn$_{2}$O$_{7}$ or Pr$_{2}$Sn$_{2}$O$_{7}$, while related members of the series with rather different properties include Tb$_{2}$Ti$_{2}$O$_{7}$ and Tb$_{2}$Sn$_{2}$O$_{7}$. The total angular momentum $\mathbf{J}=\mathbf{L}+\mathbf{S}$ is a good quantum number for rare earth elements and the electronic ground states are determined by Hund's rules.

\newcolumntype{C}{>{$}c<{$}}
\newcolumntype{L}{>{\centering $} p{2cm}<{$}}
\newcolumntype{S}{>{\centering $} p{1cm}<{$}}
\renewcommand{\arraystretch}{2}

\begin{table}[ht]
\begin{center}
\begin{tabular}{||S|C|L|S|S|C||}
\hhline{|t:======:t|}
&\textrm{[R]}&\textrm{[R}^{3+}\textrm{]}&S&L&J\\
\hhline{||------||}
\textrm{Dy}&\textrm{[Xe]} \;4f^{10}\,6s^{2}&\textrm{[Xe]} \, 4f^{9}&5/2&5&15/2\\
\textrm{Ho}&\textrm{[Xe]}\;4f^{11}\,6s^{2}&\textrm{[Xe]} \, 4f^{10}&2&6&8\\
\hhline{|b:======:b|}
\end{tabular}
\end{center}
\caption[$\;\;$ Ground state electronic configuration]{\textbf{Ground state electronic configuration} of the atom [R=\{Dy,Ho\}], the free ion [R$^{3+}$], and the corresponding quantum numbers $S, L$ and $J$}
\end{table}

The free ion Dy$^{3+}$ (resp. Ho$^{3+}$) then has a 16-fold (resp. 17-fold) degeneracy that is lifted by the surrounding crystal field; the corresponding energy levels have been estimated by neutron time-of-flight spectroscopy \cite{Rosenkranz00a}, revealing an almost pure ground state doublet.

\begin{table}[ht]
\begin{center}
\begin{tabular}{||S|C|S|C|C|C||}
\hhline{|t:======:t|}
& \textrm{GS}&g_{J}&\mu&\Delta\,(\textrm{meV})&\Delta\,(\textrm{K})\\
\hhline{||------||}
\textrm{Dy}&\left| 15/2,\pm15/2 \right\rangle&4/3&\approx10\,\mu_{B}&\sim0.033&\sim 380\\
\textrm{Ho}&\left| 8,\pm 8 \right\rangle&5/4&\approx10\,\mu_{B}&\sim0.020&\sim 240\\
\hhline{|b:======:b|}
\end{tabular}
\end{center}
\label{table:intro:CF}
\caption[$\;\;$ Crystal field level]{\textbf{Crystal field level:} The ground states $\left|J,J_{z}\right\rangle$ (GS), the Land\'e factor $g_{J}$, the estimated magnetic moment $\mu=g_{J}\,J\,\mu_{B}$ and the energy level of the first excited state $\Delta$. $\left(g_{J}=1+\frac{J(J+1)-L(L+1)+S(S+1)}{2J(J+1)}\right)$}
\end{table}

As the first excited states are separated from the doublet by an energy scale of $\Delta\approx 2-300K$ (see Table 2.), the low temperature behaviour of these materials can be approximated by classical Ising spins with a very large magnetic moment ($\approx10\,\mu_{B}$), oriented along the line connecting the centers of two neighbouring tetrahedra and forming a local set of body centered cubic axes:
\begin{gather}
\mathbf{S}_{1}\= \frac{\sigma_1}{\sqrt{3}}\,\begin{pmatrix}-1\\-1\\+1\end{pmatrix}\quad
\mathbf{S}_{2}\= \frac{\sigma_2}{\sqrt{3}}\,\begin{pmatrix}+1\\+1\\+1\end{pmatrix}\quad
\mathbf{S}_{3}\= \frac{\sigma_3}{\sqrt{3}}\,\begin{pmatrix}+1\\-1\\-1\end{pmatrix}\quad
\mathbf{S}_{4}\= \frac{\sigma_4}{\sqrt{3}}\,\begin{pmatrix}-1\\+1\\-1\end{pmatrix}.
\label{eq:intro:spins}
\end{gather}
$\sigma_i=\pm1$ is an Ising pseudo-spin with chosen convention that $\sigma_i=1$ corresponds to the moment pointing out of a \textit{down} tetrahedron (see figure~\ref{fig:intro:pyrochlore}).

\begin{figure} [ht]
\centering{\includegraphics[scale=0.7]{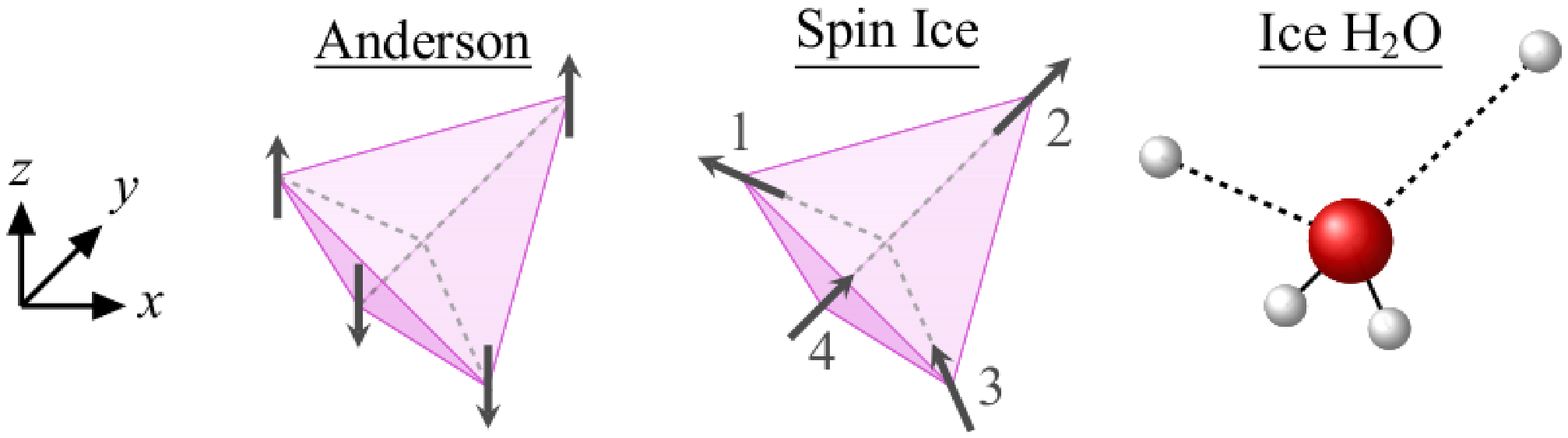}}
\caption[$\;\;$ Models equivalent to nearest neighbour spin ice]{\textbf{Models equivalent to nearest neighbour spin ice:} The different mappings between Anderson's model for antiferromagnetic spinels,  spin ice and water ice: the white spheres are the hydrogen atoms and the red one is the oxygen. A spin pointing \textit{inside} (resp. \textit{outside}) the tetrahedron corresponds to a \textit{down} (resp. \textit{up}) spin in the Anderson model and a short covalent bond (resp. long $H$-bond) for water ice.}\label{fig:intro:mapping}
\end{figure}

The physics of spin ice is captured to a good approximation by an effective model with nearest  neighbour ferromagnetic interactions between the moments which, together with the strong crystal field, gives rise to a frustrated ferromagnetic system:
\begin{eqnarray}
\mathcal{H}\=-3J_{\rm eff}\,\sum_{\langle i,j\rangle} \mathbf{S}_{i}\cdot\mathbf{S}_{j}\=J_{\textrm{eff}}\,\sum_{\langle i,j\rangle}\;\sigma_{i}\;\sigma_{j}.
\label{eq:intro:HamiltonianSI}
\end{eqnarray}
Harris \textit{\& al.}\cite{Harris97a} showed that the low energy phase space of this system is identical to the Pauling model for ice \cite{Pauling35a}, which shows extensive ground state entropy in which states satisfy the Bernal-Fowler ice rules and which has been measured, both in ice (\cite{Giauque36a}) and spin ice (\cite{Ramirez99a}). The nearest neighbour spin ice (NNSI) Hamiltonian also maps onto an Ising antiferromagnet for the pseudo spins \cite{Moessner98b,Bramwell98a} (see equation (\ref{eq:intro:HamiltonianSI})); a model first derived by Anderson to describe spinel materials \cite{Anderson56a} . The equivalence between these three nearest neighbour models is illustrated in figure ~\ref{fig:intro:mapping}.

The Bernal - Fowler ice rules requiring two magnetic moments to enter and two to leave each tetrahedron constitute a topological constraint, with the result that the Pauling states make up an ensemble of gauge invariant topological sectors with U(1) symmetry. A consequence of such constraints in this and in related systems is the development of algebraic correlations \cite{Huse03a,Hermele04a,Isakov04b,Henley05a,Henley10a}:
consider the magnetisation of a tetrahedron $\alpha$: $\mathbf{M}(\mathbf{r}_{\alpha})=\sum_{i\in \alpha} \mathbf{S}_{i}$ and define a coarse-grained field $\mathbf{M}(\mathbf{r})=\frac{1}{V}\sum_{\alpha\in V}\,\mathbf{M}(\mathbf{r}_{\alpha})$ averaged over a volume $V$ large enough to make $\mathbf{M}(\mathbf{r})$ a smoothly varying function. The proof of algebraic correlations requires a two-stage argument based on the frustrated \2 ground states.
On this manifold of states, the Gibbs free energy $G$ is purely entropic and can be expressed as a functional of $\left\{\mathbf{M}(\mathbf{r})\right\}$. For a tetrahedron respecting the ice-rules, $\mathbf{M}(\mathbf{r}_{\alpha})$ can only take 6 values proportional to $(\pm1,0,0), (0,\pm1,0)$ or $(0,0,\pm1)$. Hence computing the entropy density for a volume $V$ is  similar to the problem of a random polymer made of monomers fixed on a cubic lattice and whose extremities are separated by $\mathbf{M}(\mathbf{r})$. A large value of $\left|\mathbf{M}(\mathbf{r})\right|$ requires most of the spins/monomers to be aligned along this direction, allowing only a relatively small number of configurations.
On the other hand, zero magnetisation fixes a minimum number of degrees of freedom. Hence the Gibbs free energy and its Fourier transform are expressed to lowest order and up to a constant as
\begin{eqnarray}
G\left[\left\{\mathbf{M}(\mathbf{r})\right\}\right]&=& \frac{T}{V}\int d^{3}\mathbf{r}\,\frac{1}{2}\,K\,\left|\mathbf{M}(\mathbf{r})\right|^{2}\\
\Leftrightarrow \tilde{G}\left[\left\{\tilde{\mathbf{M}}(\mathbf{k})\right\}\right]&=&T \sum_{\mathbf{k}}\frac{1}{2}\,K\,\left|\tilde{\mathbf{M}}(\mathbf{k})\right|^{2}
\label{eq:intro:cond1}
\end{eqnarray}
which gives a temperature independent Gaussian probability $e^{-G/T}$ and $\langle \tilde{M}_{\mu}(-\mathbf{k})\tilde{M}_{\nu}(\mathbf{k})\rangle=\delta_{\mu\nu}/K$. However this argument neglects the non-local influence of the ice-rules. If we regard the spins as local fluxes of magnetisation then these constraints impose a flux conservation for each tetrahedron (2 spins \textit{in} and 2 spins \textit{out}); from a coarse-grained point of view, it is equivalent to a divergence free condition.
\begin{eqnarray}
\mathbf{\nabla}\cdot\mathbf{M}(\mathbf{r})=0\quad\Leftrightarrow \quad\mathbf{k}\cdot\tilde{\mathbf{M}}(\mathbf{k})=0
\label{eq:intro:cond2}
\end{eqnarray}
which imposes $\tilde{\mathbf{M}}(\mathbf{k})$ to be orthogonal to $\mathbf{k}$ and thus
\begin{eqnarray}
\langle \tilde{M}_{\mu}(-\mathbf{k})\tilde{M}_{\nu}(\mathbf{k})\rangle=\frac{1}{K}\left(\delta_{\mu\nu}-\frac{k_{\mu}k_{\nu}}{|\mathbf{k}|^{2}}\right)
\label{eq:intro:corrTF}
\end{eqnarray}
where the Greek indices $\mu,\nu$ are Cartesian coordinates labels. After a final Fourier transform back to real space, we obtain the desired correlations
\begin{eqnarray}
\langle{M}_{\mu}(\mathbf{0}){M}_{\nu}(\mathbf{r})\rangle\propto\frac{1}{K}\frac{\delta_{\mu\nu}-3\,\hat{r}_{\mu}\hat{r}_{\nu}}{r^{3}}\qquad\textrm{where}\quad\hat{r}_{\mu}\=\frac{r_{\mu}}{|\mathbf{r}|}
\label{eq:intro:corr}
\end{eqnarray}
This result deserves a few comments. First of all, we have lost the details of the pyrochlore lattice as soon as we used a coarse-grained field and this result can be generalised to models others than spin ice \cite{Huse03a,Hermele04a,Isakov04b,Henley05a,Henley10a}, the only hypothesis being
\begin{itemize}
\item a \textit{Coulomb phase} gauge theory imposed by equation~(\ref{eq:intro:cond1});
\item a divergence free condition, consequence of the ice-rules constraints~(\ref{eq:intro:cond2}).
\end{itemize}
The magnetisation field $\mathbf{M}(\mathbf{r})$ is then analagous to a magnetic field without monopoles which explains the specific form of the correlations similar to those generated by a \textit{dipole-dipole interaction}. One might think that the $1/r^{3}$ dependence,  in a $3d$ crystal should lead to a logarithmic divergence in the susceptibility and logarithmic peaks in scattering function. This is not, however the case as the dipolar angular dependance of the correlation function results in a cancellation of the divergent part when integrated over all space. The correlations can be observed experimentally however through the development of narrowing  \textit{pinch points} \cite{Youngblood81a}  in the diffuse scattering profile \cite{Fennell07a,Fennell09a,Kadowaki09a}.

It is clear that the NNSI model already contains extremely rich phenomenology related to the constrained Coulomb phase with $U(1)$ gauge symmetry, which is the basis for monopole physics. Most notably, single spin flip excitations create a pair of topological defects; pairs of tetrahedra with \3 and \4. The defects are expected to have an effective entropic interaction \cite{Henley10a}, but once created can move at zero energy cost. They therefore posses some but not all the characteristics of neutral magnetic monopole pairs. In order to see the emergence of true magnetic quasi-particles one has to go beyond the NNSI through the inclusion of dipolar interactions, as is considered in the next section.

\section{The dipolar spin ice model}
\label{intro:DSI}

Dipolar interactions are of particular importance \cite{Siddharthan99a,Hertog00a} for the rare earth ions Ho$^{3+}$ and Dy$^{3+}$ at the heart of spin ice physics: as the exchange coupling between the ions is due to $4f$ electrons, buried behind $5s$ and $5p$ layers\footnote{The $6s$ layer is empty because Ho$^{3+}$ and Dy$^{3+}$ lost three electrons.}, it is very weak ($\sim$ 1K) in comparison with other ferromagnets (for example $J \sim 10^{2}$ K for metallic iron). The magnetic moments in spin ice being very large,  the dipolar interactions, that are negligible on short length scale for iron, are on the same energy scale as the exchange here. A quantitative description has therefore to take both interactions into account
while to an excellent approximation moments can be described by Ising degrees of freedom orientated along the body centers of the tetrahedtra.
This separation of scales leads to the  dipolar spin ice (DSI) Hamiltonian \cite{Hertog00a}
\begin{eqnarray}
\mathcal{H}\;=\;-J\,\sum_{\left<i,j\right>}\mathbf{S}_{i}\cdot\mathbf{S}_{j}\;+\;
D\,r_{nn}^{3}\sum_{i > j}\left[\frac{\mathbf{S}_{i}\cdot\mathbf{S}_{j}}{\left|\mathbf{r}_{ij}\right|^{3}}- \frac{3\left(\mathbf{S}_{i}\cdot\mathbf{r}_{ij}\right)\left(\mathbf{S}_{j}\cdot\mathbf{r}_{ij}\right)}{\left|\mathbf{r}_{ij}\right|^{5}}\right],
\label{eq:intro:HamiltonianDSI}
\end{eqnarray}
where $J, D$ and $r_{nn}\approx3.5$ \AA$\,$ are respectively an antiferromagnetic exchange coupling, the dipole-dipole coupling and the nearest neighbour distance between rare earth ions, which has been shown to provide a comprehensive quantitative description of spin ice materials.  Here, $J$ is a free parameter, while $D$ is defined through the dipolar interaction
\begin{eqnarray}
D=\frac{\mu_{0}\;\mu^{2}}{4\pi\;r_{nn}^{3}}\approx 1.4 K,
\label{eq:intro:D}
\end{eqnarray}
for Dy$_2$Ti$_2$O$_7$. A numerical estimate of $J$ has been obtained by comparison between Monte Carlo simulations and experimental data for the specific heat in Dy$_{2}$Ti$_{2}$O$_{7}$ \cite{Hertog00a} or of neutron scattering on Ho$_{2}$Ti$_{2}$O$_{7}$ \cite{Bramwell01a} and  Ho$_{2}$Sn$_{2}$O$_{7}$ \cite{Kadowaki02a}. Uncertainties in the estimated values for the Stanates remain high;  around 50\% for Ho$_{2}$Sn$_{2}$O$_{7}$ because of the unavailability of single crystals. 
The dipolar contribution turns out to be indispensable for understanding certain magnetic field induced transitions \cite{Melko04a,Yoshida04a,Higashinaka05a,Ruff05a}  as well as the quantitative feature of magnetic monopole excitations \cite{Jaubert09a}. However, a remarkable property of this Hamiltonian is that the extensive degeneracy of an effective nearest neighbour model is almost maintained, as the nearest neighbour contribution to the dipolar interaction is the same for all Pauling states and the long ranged part of the interaction is almost perfectly screened\cite{Gingras01a,Isakov05a}.  The dipolar interaction between neighbouring spins 1 and 2 defined in figure~\ref{fig:intro:mapping} is given by:
\begin{eqnarray}
&&D\,r_{nn}^{3}\left[\frac{\mathbf{S}_{1}\cdot\mathbf{S}_{2}}{\left|\mathbf{r}_{12}\right|^{3}}- \frac{3\left(\mathbf{S}_{1}\cdot\mathbf{r}_{12}\right)\left(\mathbf{S}_{2}\cdot\mathbf{r}_{12}\right)}{\left|\mathbf{r}_{12}\right|^{5}}\right]\nonumber\\
&=&\frac{5D}{3}\sigma_{1}\sigma_{2}.
\end{eqnarray}
By symmetry, the result is the same for all pairs of spins in a tetrahedron. Hence the NN term of the dipolar interaction is equivalent to an exchange term and the nearest neighbour spin ice (NNSI) Hamiltonian is recovered as follows
\begin{eqnarray}
\mathcal{H}\= -3J_{\rm eff}\,\sum_{\langle i,j\rangle}\; \mathbf{S}_{i}\cdot\mathbf{S}_{j}\=J_{\textrm{eff}}\,\sum_{\langle i,j\rangle}\;\sigma_{i}\;\sigma_{j}\quad
\textrm{where}\quad J_{\textrm{eff}}\=D_{nn}+J_{nn}\=\frac{5D}{3}+\frac{J}{3}
\label{eq:intro:trueHamiltonianSI}
\end{eqnarray}
Values for the spin ice materials is summarised in table 3. below.

\begin{figure} [ht]
\centering{\includegraphics[scale=0.7]{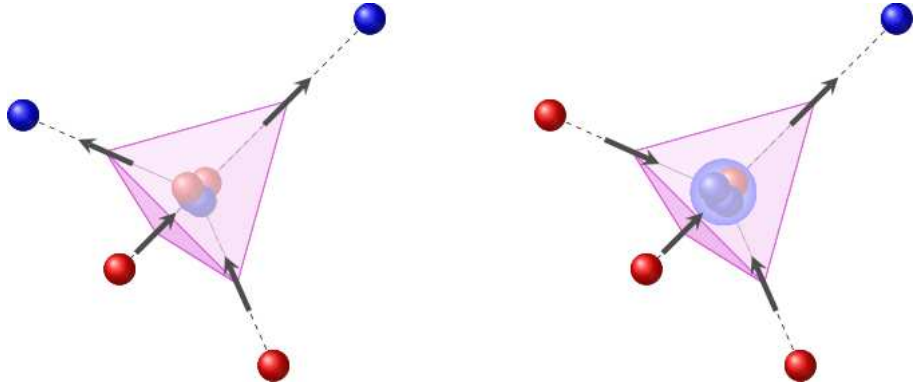}}
\caption[$\;\;$ Dumbbell model]{\textbf{Dumbbell model:} Each magnetic dipole is seen as two charges sitting on the vertices of the diamond lattice (dashed lines). \2 is a vacuum whereas \3 is a positive charge.}
\label{fig:intro:dumbbell}
\end{figure}

\begin{table}[ht]
\begin{center}
\begin{tabular}{||C|C|C|C|C|C||}
\hhline{|t:======:t|}
&D&D_{nn}&J&J_{nn}&J_{\textrm{eff}}\\
\hhline{||------||}
\textrm{Dy~\protect \cite{Hertog00a}}&1.41&2.35&-3.72&-1.24& \approx 1.1\\
\textrm{HoTi~\protect \cite{Bramwell01a}}&1.41&2.35&-1.65&-0.52& \approx 1.8\\
\textrm{HoSn~\protect \cite{Kadowaki02a}}&1.41&2.35&\sim 1.0&\sim 0.33&\sim 2.7\\
\hhline{|b:======:b|}
\end{tabular}
\end{center}
\label{table:intro:JDJeff}
\caption[$\;\;$ Nearest neighbour interactions]{\textbf{Nearest neighbour interactions:} Exchange, dipolar and effective anti-ferromagnetic couplings in Kelvin.}
\end{table}
The origin of this approximate screening, the so called  projective equivalence, \cite{Isakov05a} can be seen physically by considering a further abstraction of the DSI to the dumbbell model \cite{Castelnovo08a}, which highlights the remarkable symmetry properties of dipoles on the pyrochlore lattice. In this model the point dipoles of the DSI are extended into magetically charged dumbbells, as shown in figure (\ref{fig:intro:dumbbell}),  whose ends sit on the centers of the tetrahedra,  forming a diamond lattice of magnetically neutral sites for all Pauling states. Projective equivalence is clearly exact for this model and the error incurred in DSI, which turn out to be  quadrupolar corrections only\cite{Isakov05a} is due to the difference between point and extended dipoles.

\section{Magnetic monopoles and classical "Dirac strings"}

\begin{figure} [ht]
\centering\includegraphics[scale=0.8]{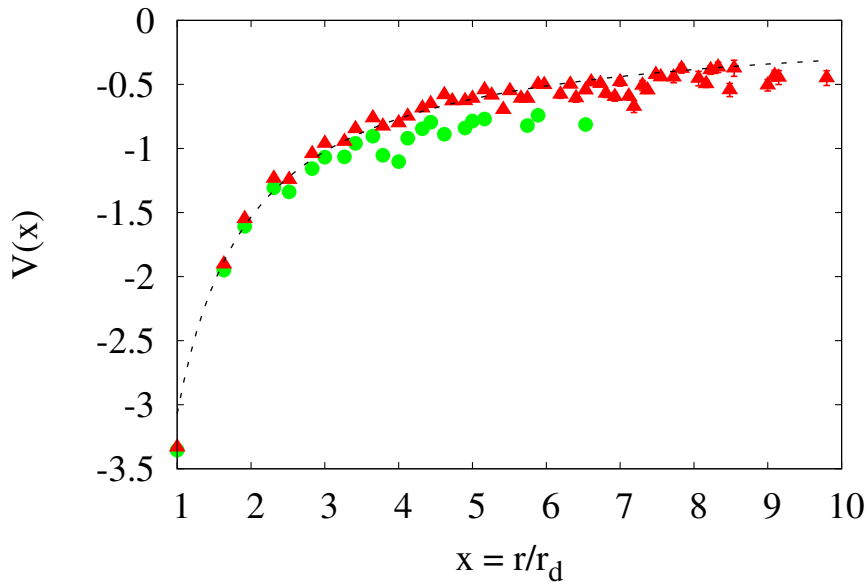}
\caption[$\;\;$ Coulomb potential]{\textbf{Coulomb potential:} Effective interaction (in Kelvin) between a pair of monopoles in the dipolar spin ice model as a function of the distance between them in units of nearest neighbour distance between monopoles. The system sizes are $4\times4\times4$ unit cells, \ie 1024 spins (green crosses) and $8\times8\times8$ unit cells, \ie 8192 spins (red dots). The solid line is the Coulomb potential expected from the dumbbell model $V(x)=V_{\textrm{min}}/x$. As we are on the diamond lattice, the smallest possible distance between monopoles is $x=1$. The data are the average of over more than 100 samples.}
\label{fig:mono:coulomb}
\end{figure}

Within the dumbbell model, the ensemble of Pauling states provide a quasi-particle vacuum for neutral pairs of topological defects. Once created, the \3 and \4 defects break magnetic charge neutrality on each duel lattice, so that, when separated, the defects should interact via Coulombs law. Hence, one can see by construction that fractionalization  of the dipolar dumbbells into deconfined magnetic Coulomb charges occurs when the ice rules are broken: in this sense these quasi-particles are magnetic monopoles. To show that, thanks to projective equivalence, the same is true to an excellent approximation in the DSI, we have followed \cite{Castelnovo08a} and simulated a DSI system.
The long range interactions have been treated using the Ewald method \cite{Deleeuw80a,Frenkel02a,Melko04a}, where the slowly converging sum of all dipolar interactions between spins is replaced by two strongly convergent series. Periodic boundary conditions ensures correct convergence.
Starting with a random \2 configuration, we can create and force the diffusion of a single pair of monopoles by flipping spins and then compute the energy of the system for each configuration. After averaging over a large number of initial microstates and paths of diffusion, we obtain the potential of interaction between two monopoles of opposite charges shown in figure \ref{fig:mono:coulomb}. For a distance $r=x\,r_{d}$ where $r_{nn}$ is the distance between two vertices on the diamond lattice (see figure~\ref{fig:intro:pyrochlore}) , the Coulomb potential can be written \cite{Castelnovo08a}
\begin{eqnarray}
V(r)\=-\frac{\mu_{0}}{4\pi}\frac{Q^{2}}{r}\=\frac{V_{\textrm{min}}}{x},\quad\textrm{where}\quad Q=\frac{2\mu}{r_{d}},
\label{eq:mono:V1}
\end{eqnarray}
and where
\begin{eqnarray}
V_{\textrm{min}}\=-\frac{\mu_{0}}{4\pi}\frac{4\,\mu^{2}}{\frac{3}{2}r_{nn}^{2}}\frac{1}{\sqrt{\frac{3}{2}}\,r_{nn}}\=-\frac{8}{3}\,\sqrt{\frac{2}{3}}\,D\approx -3.07 \textrm{ K},
\label{eq:mono:V2}
\end{eqnarray}
with, from equation~(\ref{eq:intro:D}), $D\approx 1.41$ K. $V_{\textrm{min}}$ is the energy gained, with respect to the vacuum, by the creation of a nearest neighbour pair of monopoles. This potential $V$ for the dumbbell model is compared with our numerical results for DSI in figure~\ref{fig:mono:coulomb}.
As we have periodic boundaries, the pair of monopoles under consideration will also feel the presence of its  \textit{image charges} separated by at least one system size. While this is convenient for simulating a real system (as is done later in this paper), it leads to a finite size bias to the Coulomb energy here, where we consider a single pair of monopoles. This bias remains negligible for small $x$ but can be observed at larger $x$ and scales away with system size. This is illustrated in~\ref{fig:mono:coulomb} for two system sizes $L=4$ (green crosses) and $L=8$ (red dots). For the larger system size the $1/x$ behaviour is clearly observed and the data respects the energy scale fixed by $V_{\textrm{min}}$ to a good approximation. The data plotted here are  averaged over at least $100$ simulations, so that the small differences between our results and Coulomb's law reflect the limits of accuracy of the dumbbell model, which are of order $1/r^{5}$. The only fitting parameter in the figure is the reference energy scale for the 
DSI, which translates the numerical data along the $y-$axis.
Similar results are obtained in reference \cite{Castelnovo08a}. However while they computed the energy of separation between monopoles for a single configuration, we have performed a statistical average, which produces some statistical noise, but confirms that this is a general property of the ensemble of constrained Pauling states.
\begin{figure}[ht]
\centering\includegraphics[scale=0.8]{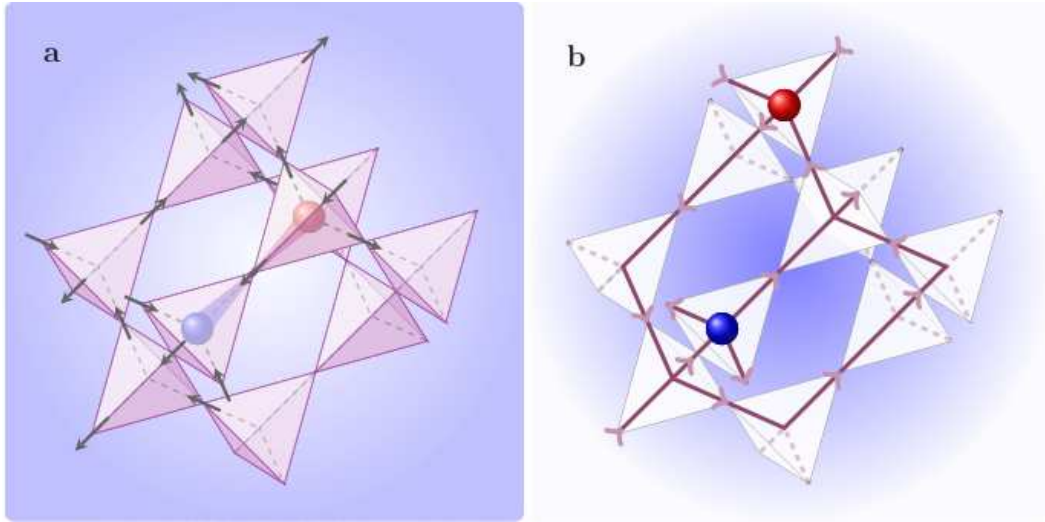}
\caption[$\;\;$ Emergence of monopoles]{\textbf{Emergence of monopoles} a) The magnetic ions (Ho$^{3+}$ or Dy$^{3+}$) lie on the sites of the pyrochlore lattice and are constrained to the bonds of the diamond lattice (dashed lines). Local topological excitations \3 or \4 correspond to magnetic monopoles with positive (blue sphere) or negative (red sphere) charges respectively. b) The diamond lattice provides the skeleton for the network of Dirac strings with the position of the monopoles restricted to the vertices. The orientation of the Dirac strings shows the direction of the local field lines in $\mathbf{H}$.}
\label{fig:mono:mono}
\end{figure}\\
Table 4. summarises the main parallels between the dumbbell and DSI models.
\renewcommand{\arraystretch}{1.2}
\renewcommand{\baselinestretch}{1.2}
\newcolumntype{D}{>{}c<{}}
\begin{table}[ht]
\begin{center}
\begin{tabular}{||D|D||}
\hhline{|t:==:t|}
\textbf{Dipolar Spin Ice} & \textbf{Dumbbell / Monopoles}\\
\hhline{||==||}
\2 ground state& vacuum\\
\hhline{||--||}
\3 defects (local excitations) & quasi-particles (monopoles)\\
\hhline{||--||}
dipolar interactions between& Coulomb interactions between\\
magnetic dipoles & magnetic monopoles\\
\hhline{||--||}
pyrochlore lattice & diamond lattice\\
\hhline{||--||}
canonical ensemble & grand canonical ensemble\\
\hhline{|b:==:b|}
\end{tabular}
\end{center}
\renewcommand{\arraystretch}{1.}
\renewcommand{\baselinestretch}{1.}
\caption[$\;\;$ Mapping from the dipolar spin ice to the dumbbell model]{\textbf{Mapping from the dipolar spin ice to the dumbbell model}}
\end{table}

The theoretical basis for magnetic monopoles in the DSI model is thus clearly established. We particularly stress that the scale of the Coulomb interaction is set by \textit{magnetic} constants: the permeability of free space, $\mu_0$, and the nearest neighbour distance on the diamond lattice. Hence the quasi-particles behave quantitatively as  \textit{magnetically} charged particles. However, two fundamental differences still exists between these particles and Dirac monopoles: firstly they correspond to divergences in the magnetic intensity $\mathbf{H}$, or magnetic moment $\mathbf{M}$, rather than in the magnetic induction, so that $\nabla\cdot\mathbf{B}=\mathbf{\nabla}\cdot(\mathbf{H}+\mathbf{M})=0$ and they do not require a modification of Maxwell's equations. Secondly, their magnetic charge is not quantified. The magnetic charge on a Dirac monopole is quantified through a quantum conjugation relation with the fundamental electronic charge \cite{Jackson99a,Dirac31a}. This is clearly not the case in spin ice and in fact the charge can be continuously varied by modifying the nearest neighbour distance between spins by applying external pressure \cite{Castelnovo08a}. 
However, on all length scales above the atomic scale, a \3 defect appears to be a local sink in $\mathbf{M}$ and therefore a source of field lines in $\mathbf{H}$. Furthermore we can assign to them a positive or negative charge when immersed in a magnetic field \cite{Ryzhkin05a} that is conserved through the creation and annihilation of neutral monopole pairs. As the Coulomb potential is not confining\footnote{having a Coulomb gas on a \textit{lattice} prevents the divergence for $r \rightarrow 0^{+}$. The energy of deconfinement is finite and equals to $\left|V_{\textrm{min}}\right|$ on average.} in $3d$, these quasi-particles provide an example of fractionalisation in high dimensions: the point dipoles of the DSI appear to be separated into free poles of opposite sign that are free to wander independently through the system. They are therefor classical analogues of Dirac monopoles and spin ice provides the first $3d$ experimental realisation of deconfined magnetic monopoles.

In the theory of Dirac, monopoles of a neutral pair of charges are connected by a tensionless Dirac string of overturned dipoles, taken in the the continuum limit. The string, which corresponds to nodes in the wave function for the pair of particles, should be unobservable. Classical analogues of Dirac strings exist in DSI and in the dumbbell model \cite{Castelnovo08a}, which correspond to the magnetic moments in the constrained \2 states. That is, by creating and separating a pair of monopoles one flips a series of moments, defining one possible Dirac string connecting them. Starting from a disordered \2 state, the path taken to create the monopoles and place them at positions $A$ and $B$ is not unique and the arbitrariness provides a classical analogue of the quantum uncertainty of  Dirac monopole trajectoires in space. Further to this analogy, within the dumbbell model, as long as the dumbbells are taken to have point magnetic charges at each end (the dumbbells can also be thought of as infinitesimally thin magnetic needles, touching at the centres of the tetrahedra), the magnetic flux associate with each dipole is confined to infinitely narrow elements joining the charges on each dumbbell,  so that the network of classical Dirac strings is also invisible. This of course is a mathematical abstraction; in the DSI, we have point dipoles on the vertices of the tetrahedra and because of this difference (because of corrections to projective equivalence in fact), the string network {\it is} observable. Indeed, recent experiments have reported the observation of "Dirac strings"\cite{Fennell09a,Morris09a}. As far as dynamics are concerned, the Dirac string network plays a crucial role: moving a monopole involves a spin flip, so that the quasi-particles leave a wake of re-arranged links in the network of Dirac strings as they move, with \3 and \4 defects moving in opposite directions over the network (see figure \ref{fig:mono:mono}). This directionality makes the dynamics highly constrained compared with the diffusing electric charge of an equivalent electronic Coulomb gas. At equilibrium and in zero field this constraint simply renormalizes the diffusion constant but in a rapid quench the constraints lead to glassy behaviour \cite{Castelnovo10b} reminiscent of other kinetically constrained models\cite{Mayer06a}. We will return to the dynamics in finite field in the conclusion.

Given the accessibility of these magnetic quasi-particles, the development of an experimental signature is of vital importance and interest and has recently given rise to considerable experimental activity\cite{Fennell09a,Morris09a,Kadowaki09a,Giblin10a}.  In particular, recent experiments have measured the charge carried on free deconfined particles via the magnetic Wien effect \cite{Onsager34a,Bramwell09a}. Here, we will concentrate on magnetic relaxation,
available from measurements of magnetic response in weak ac magnetic fields \cite{Matsuhira00a, Matsuhira01a,Ehlers03a,Snyder04b}. In the following section we will show how these data can be used as a direct signature for diffusive monopole dynamics.

\section{Dynamics in spin ice}

\begin{figure} [ht]
\centering\includegraphics[scale=0.7]{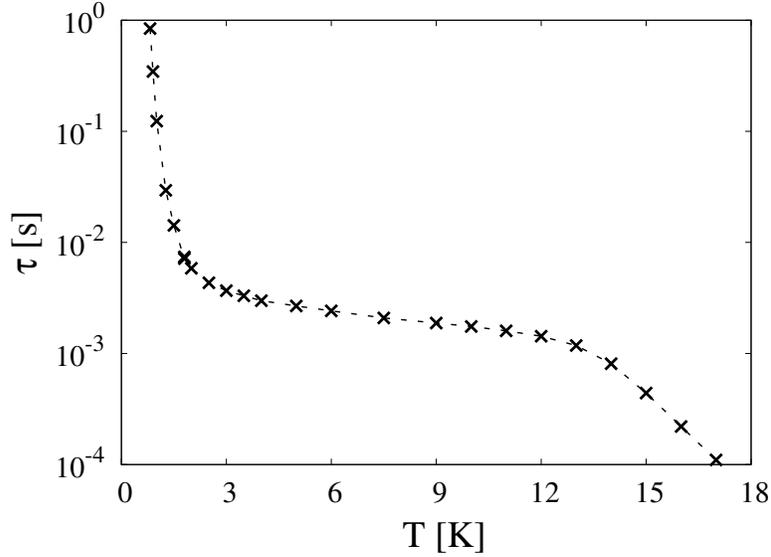}
\caption[$\;\;$ Experimental relaxation time]{\textbf{Experimental relaxation time $\mathbf{\tau}$} for Dy$_{2}$Ti$_{2}$O$_{7}$ vs temperature $T$ from Snyder \textit{\& al.} \protect \cite{Snyder04b}, courtesy of Peter Schiffer. The starting point of this work has been to shed new light on this result. Above 12 K the spins are not exactly Ising anymore. The quasi-plateau between 2 and 12 K is divided in two parts: the ``high temperature'' of the spin ice region above 4 K where one needs to consider double defects 4 \textit{in} or 4 \textit{out} and a large monopole density, and the ``low temperature'' regime below 6 K where our monopole model should hold. Below $\sim 2$ K, the system becomes frozen.}
\label{fig:SF:Snyder}
\end{figure}

In figure (\ref{fig:SF:Snyder}) we show the evolution of the magnetic relaxation time scale for Dy$_{2}$Ti$_{2}$O$_{7}$ measured from bulk {\it a.c.} susceptibility measurement \cite{Snyder04b}. The data show three distinct features: a high temperature regime above $12$ K, which, has been associated with thermal excitations of higher crystal field levels \cite{Snyder01a,Ehlers03a,Ehlers04a}, an intermediate temperature plateau regime between $3$ and $10 K$ and a regime with rapidly increasing time scale below $3$ K. Qualitatively  similar behaviour has been observed in Holmium Titanate\cite{Clancy09a}. We will use this data to illustrate that the low and intermediate temperature dynamics of spin ice can be interpreted in terms of the stochastic dynamics of monopole quasi-particles. Given the separation of energy scales in spin ice materials we follow the assumption that the dynamics in the low temperature and the plateau regions are controlled by quantum tunneling events \cite{Ehlers03a} taking the magnetic moments directly from one Ising position to another through the existence of non-zero off-diagonal components of the dipolar interaction from neighbouring spins, giving a small mixing of the magnetic states. Within this picture an excitation requires thermal assistance to make an Ising spin flip, but the time scale for the flip should be given by rate constant $\tau_0$, which is essentially temperature independent. The high temperature limit of this model is then an Ising paramagnetic system evolving with temperature independent rate constant, giving a plateau in the evolution of the relaxation time scale. The deviation from the plateaux above $10$ K is due to thermal mixing of the crystal field levels \cite{Snyder01a,Ehlers03a,Sutter07a,Shi07a,Lago07a,Kitagawa08a} and will not be discussed further here.

From figure (\ref{fig:SF:Snyder}) the tunneling rate, $\tau_0$ for Dy$_{2}$Ti$_{2}$O$_{7}$, appears to be around $10^{-3}$ s , while for Ho$_{2}$Ti$_{2}$O$_{7}$, it is on the scale of $\tau_{o}\approx 10^{-8}$ s \cite{Clancy09a}. These extremely slow, yet exponentially varying microscopic time scales from one compound to another, seem consistent with a quantum tunneling process exponentially dependent on the energy levels of the rare-earth ions CEF.
Unfortunately, the complexity of these CEF prevents quantitative estimates for the experimental inversion rate and the time scale $\tau_{o}$ remains a fitting parameter in our theory.
The quantum tunneling on these microscopic time scales should lead to stochastic dynamics on all length scales and if a single micropscopic process dominates the dynamics, one should be able to model the relaxation using an Arhenius law. For a more complex system, for example one with long range interactions, one should be able to use the Metropolis dynamics of a Monte Carlo simulation to model the evolution of the system. These techniques and the validity of the model, with temperature independent $\tau_0$ are discussed in detail in the next section.

\subsection{Arrhenius law and deconfined quasi-particles}

\begin{figure}
\centering{\includegraphics[scale=0.65]{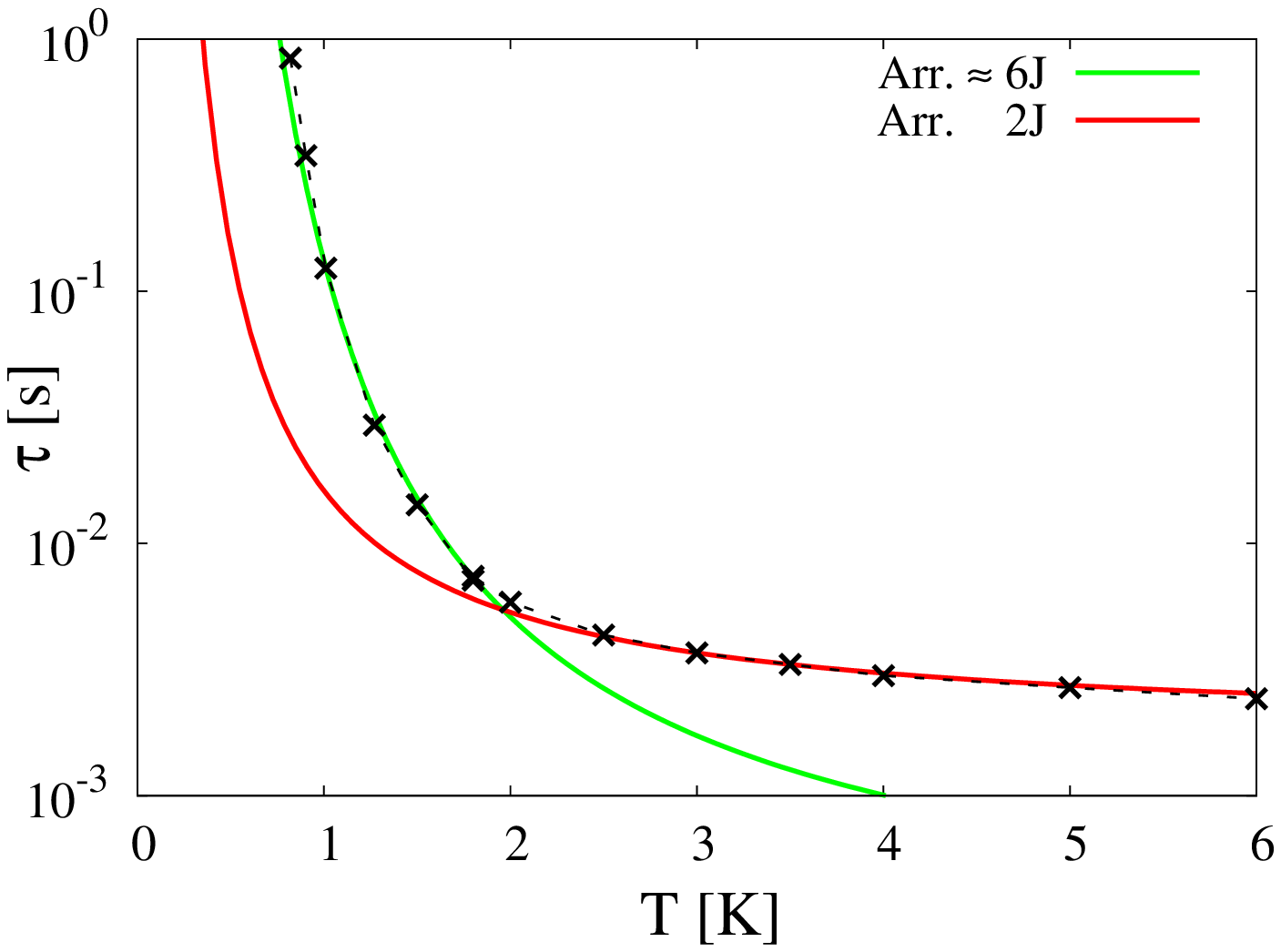}}
\centering{\includegraphics[scale=0.65]{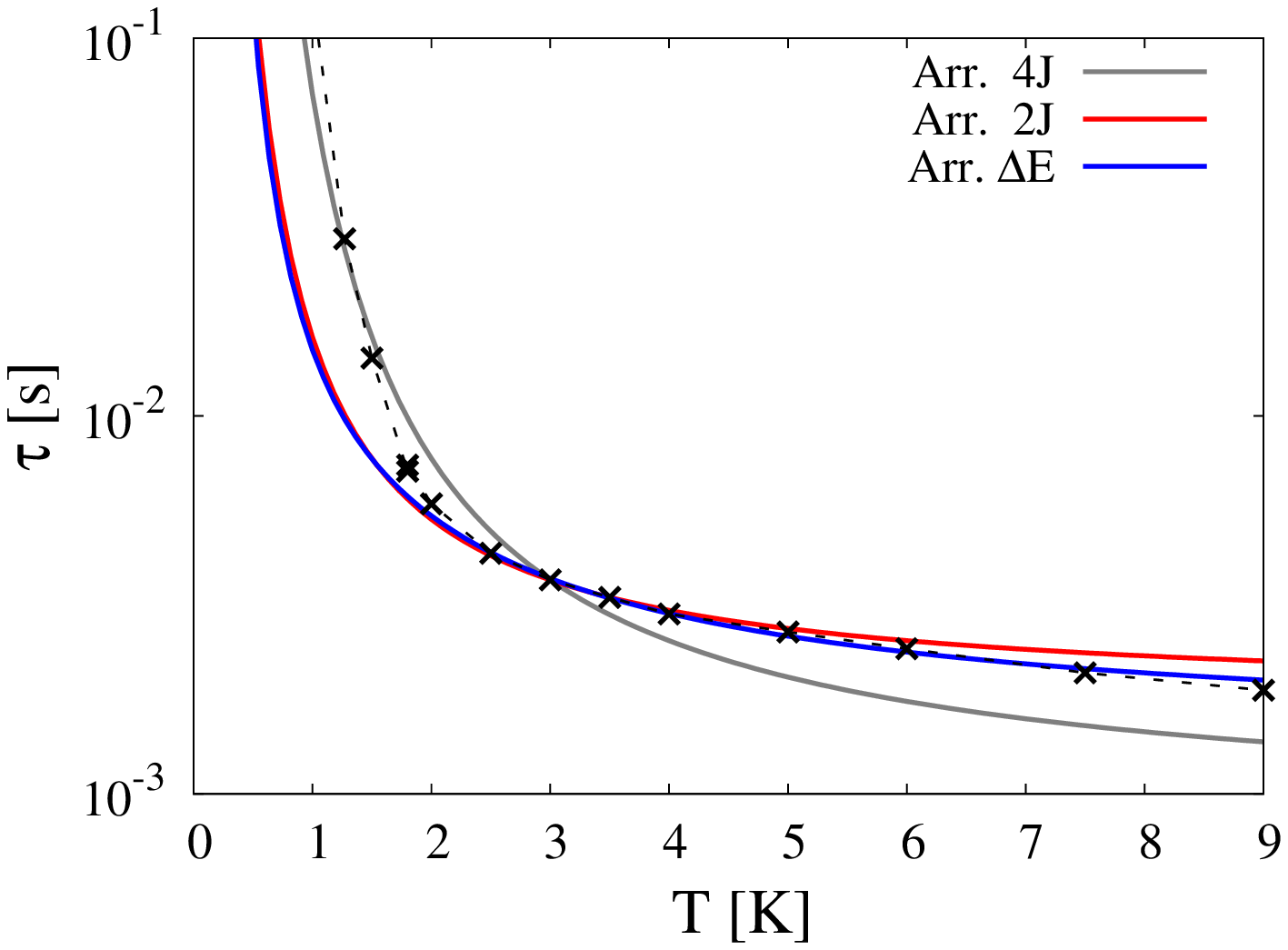}}
\caption[$\;\;$ Deconfined monopoles]{\textbf{Deconfined monopoles:} The experimental data of Snyder \textit{\& al.} \protect \cite{Snyder04b} ($\mathbf{\times}$ and dashed line) are compared with different Arrhenius functions. \textit{Top:} The quasi-plateau region is in quantitative agreement with a thermally activated process with energy barrier $E_{p}=2\,J_{\textrm{eff}}$ (red line) whereas the spin freezing is well reproduced with $E_{p}\approx 6\,J_{\textrm{eff}}$ (green line), but no unique function can fit the whole temperature window. \textit{Bottom:} The characteristic excitation is the creation of a unique defect ($E_{p}=2\,J_{\textrm{eff}}$, red line) rather than a single spin flip ($E_{p}=4\,J_{\textrm{eff}}$, grey line). The fit is improved at higher temperature if we include the energy scales due to double defects in a modified Arrhenius law (blue line).}
\label{fig:mono:Arr}
\end{figure}

The first quantitative analysis of spin ice dynamics \cite{Matsuhira00a} relates to Ho$_{2}$Sn$_{2}$O$_{7}$ and Ho$_{2}$Ti$_{2}$O$_{7}$. Here the relaxation time, $\tau(T)$ was fitted with an Arrhenius law in the freezing regime ($T\approx 1$ K) with characteristic energy barrier, $E_{f}=19.6$ K and 27.5 K respectively. These numbers already suggest that a simple Arrhenius law cannot give a complete description of the dynamics, as they do not correspond to any of the energy scales appearing in the DSI:
\begin{itemize}
\item $\sim 200$ K between the ground state doublet and the first excited crystal field levels;
\item $\sim 4 J_{\textrm{eff}}= 7.2$ K for Ho$_{2}$Ti$_{2}$O$_{7}$ and 4.4 K for Dy$_{2}$Ti$_{2}$O$_{7}$, due to single spin flips within the NNSI model;
\item $\sim 3$ K for the limit of infinite separation between monopoles (see figure~\ref{fig:mono:coulomb}).
\end{itemize}
Quantitative studies of Dy$_{2}$Ti$_{2}$O$_{7}$ rule out such  a simple thermally activated process \cite{Matsuhira01a,Snyder04b}, as shown in figure~\ref{fig:mono:Arr}. Here, the data of Snyder \textit{\& al.}  are fitted in the low temperature regime with an energy barrier  of $E_{f}=6J_{\rm eff}\sim 6.6$ K, with $J_{\textrm{eff}}=1.11$ K, the value estimated for Dy$_{2}$Ti$_{2}$O$_{7}$ \cite{Hertog00a}. The Arrhenius scaling is quantitatively good below 2 K, but fails completely to reproduce the quasi-plateau region. As we have argued above, if the tunneling picture is correct,
the plateau region, through its simplicity, offers an anchor point for any fit and any procedure that does not fit here cannot contain the physics of the problem. 

Rather than fitting the low temperature freezing region one should therefore start by fitting the data through the plateau region. By doing this we see the first evidence for deconfined quasi-particle excitations.
 Following the idea that the quasi-plateau is in fact the \textit{tail} of an exponential, thermally activated process, we have plotted in figure~\ref{fig:mono:Arr} different Arrhenius laws $\tau_{0}\,\exp(E_{p}/T)$ where the energy barrier is varied. The time scale $\tau_{0}$ is fixed by scaling to the experimental time at 4 K; a temperature at which we expect the  density of double defects (4 \textit{in} and 4 \textit{out}) to be negligible, leading to a direct relationship between spin flips and monopole pair creation. From the figure one can see the main result of this section: the experimental data are poorly reproduced by an energy barrier $E_{p}=4\,J_{\textrm{eff}}$, the cost of a single spin flip within the NNSI model, but are quantitatively well fitted with $E_{p}=2\,J_{\textrm{eff}}$ over the region between 2.5 K and 5 K (see the lower panel of figure~\ref{fig:mono:Arr}). This means that the lowest excitation responsible for the dynamics of spin ice materials is \textit{not} the energy cost of a single spin flip, but only half of that \ie the energy cost of a single topological defect. Within this effective temperature range, the NNSI model seems sufficient to capture the physics, as suggested by polarized neutron scattering \cite{Chang10a}. Once a pair of defects is created, they can propagate freely and separately in the system, leading to magnetic relaxation from independent (deconfined) objects. 

As shown in figure~\ref{fig:mono:Arr}, some corrections to Arrhenious behaviour develop above 5 K, which we show in the next section, is due to the excitation of 4 \textit{out} and  4 \textit{in} double defects, by constructing a variant of the Arrhenius law, including all relevant energy scales for single spin flips within the context of the NNSI model.

\subsection{Multi-energy Arrhenius law}

We use an independent tetrahedron picture, assuming that the probability of finding a  tetrahedron in a certain state is simply given by the Boltzmann weights of the two tetrahedra involved in any spin flip. We label the different configurations \2, \3 (or \4) and 4 \textit{in} (or 4 \textit{out}) 2, 3 and 4 respectively. In table IV.2 we show a list of all energy changes resulting from a single spin flip shared between two tetrahedra, from which we can calculate the mean energy barrier $\Delta E$.

\renewcommand{\arraystretch}{1.1}
\renewcommand{\baselinestretch}{1.2}
\newcolumntype{A}{>{$}c<{$}}
\newcolumntype{M}{>{\centering $} p{2.5cm}<{$}}
\begin{table}[ht]
\begin{center}
\begin{tabular}{||M|M|M|A||}
\hhline{|t:====:t|}
\textrm{Initial}&\textrm{Final}&\delta E &\textrm{Probability}\\
\hhline{||====||}
2|2 & 3|3 & +\;4\, J_{\textrm{eff}} & 1\\
\hhline{||----||}
2|3 & 3|2 &          0           & 3/4\\
       & 3|4 & +\;8\, J_{\textrm{eff}} & 1/4\\
\hhline{||----||}
2|4 & 3|3 &  -\;4\, J_{\textrm{eff}} & 1\\
\hhline{||----||}
3|3 & 2|2 &  -\;4\, J_{\textrm{eff}} & 9/16\\
       & 2|4 & +\;4\, J_{\textrm{eff}} & 6/16\\
       & 4|4 & +12\, J_{\textrm{eff}} & 1/16\\
\hhline{||----||}
3|4 & 4|3 &           0          & 1/4\\
       & 2|3 &  -\;8\, J_{\textrm{eff}} & 3/4\\
\hhline{||----||}
4|4 & 3|3 &  -12\, J_{\textrm{eff}} & 1\\
\hhline{|b:====:b|}
\end{tabular}
\end{center}
\label{table:mono:Arr}
\renewcommand{\arraystretch}{1.}
\renewcommand{\baselinestretch}{1.}
\caption[$\;\;$ Energy scales resulting from a single spin flip]{\textbf{Energy scales resulting from a single spin flip:} Let us explain this table using the second row as an example. It means that if we flip a spin that initially belongs to a tetrahedron \2 and a tetrahedron \3, then there is a probability of $3/4$ simply to inverse the position of the defect ($3|2$) at no energy cost, and a probability of $1/4$ to create an additional pair of defects ($3|4$) that cost an energy $\delta E=+8\,J_{\textrm{eff}}$. The probabilities are a statistical average, whether the spin under consideration is one of the three \textit{in} spins of the \3 tetrahedron ($\delta E=0$), or the fourth \textit{out} spin ($\delta E=8\,J_{\textrm{eff}}$).}
\end{table}
There are two points worth noting: the configurations ($2|3$) and ($3|2$) are different, which gives a factor of 2 for their Boltzmann weight. More importantly, a single spin flip that actually creates a pair of defects will effectively only cost one half of $\delta E$, because of the fractionalisation of the excitations, but the process ($3|3$)$\rightarrow$($2|4$) really costs $\delta E=+4\, J_{\textrm{eff}}$ as it simply moves a charge. We find $\Delta E$:
\begin{eqnarray}
\Delta E(T)&=&\frac{2\left(6\,e^{2\beta J_{\textrm{eff}}}\right)^{2}\+4\left(2\,\frac{1}{4}\,6\,e^{2\beta J_{\textrm{eff}}}\;8\right)\+6\left(\frac{1}{16}\,8^{2}\right)\+4\left(\frac{6}{16}\,8^{2}\right)}
{\left(6\,e^{2\beta J_{\textrm{eff}}}\right)^{2}\+\left(2\,\frac{1}{4}\,6\,e^{2\beta J_{\textrm{eff}}}\;8\right)\+\left(\frac{1}{16}\,8^{2}\right)\+\left(\frac{6}{16}\,8^{2}\right)}\;J_{\textrm{eff}}\nonumber\\
&=&3\;\frac{6\,e^{4\beta J_{\textrm{eff}}}\+8\,e^{2\beta J_{\textrm{eff}}}\+10}{9\,e^{4\beta J_{\textrm{eff}}}\+6\,e^{2\beta J_{\textrm{eff}}}\+7}\;J_{\textrm{eff}}\xrightarrow[T\rightarrow0]{}\;2\,J_{\textrm{eff}},
\label{eq:mono:Arr}
\end{eqnarray}
from which we recover the creation energy of a single defect in the low temperature limit. The modified Arrhenius law $\tau_{0}\,\exp(\Delta E(T)/T)$ is plotted in figure~\ref{fig:mono:Arr} (blue curve). It is almost identical to the thermally activated process with $E_{p}=2\,J_{\textrm{eff}}$ (red line) below 4 K, but differs at higher temperature and follows the experimental data more closely up to $\approx 8$ K. Quantitative differences develop above this limit as one expects: above this temperature the thermally activated processes of the high temperature regime should make their presence felt. Using the function 
$\tau_{0}\,\exp(\Delta E/T)$ clearly extends the range over which this phenomenological model can describe the data and we believe that it convincingly confirms the validity of the quantum tunneling model to describe the dynamics up to $8-10K$. We believe that this phenomenological approach provides a very strong argument in favour of  magnetic relaxation assisted by the thermally activated creation of deconfined quasi-particles in spin ice and hence in favour of \textbf{fractionalisation}  of dipoles into free effective monopoles !

However as shown in ~\ref{fig:mono:Arr}, any Arrhenius function ultimately fails below 2-3 K, underestimating the time scale in the low temperature region. Along  the quasi-plateau the density of monopoles is high enough for Debye screening to occur with the result that effective models based on the NNSI provide a convincing description. As the temperature is lowered, their density is reduced and a quantitative description can only be achieved by taking explicitly into account the long range Coulomb interactions. In this case, simple Arrhenius behaviour is excluded and we must revert to numerical simulation.

\section{Spin freezing}

We have simulated the dynamics of both the DSI model and a Coulomb gas of magnetic monopoles in the grand canonical ensemble occupying the sites of the diamond lattice, using a Metropolis Monte Carlo algorithm. The Coulomb particles are excitations out of the quasi-particle vacuum provided by the dumbbell model and their dynamics is constrained by the underlying network of Dirac strings. In both cases we treat the long range interactions, either dipolar or Coulombic using the  Ewald method \cite{Deleeuw80a,Frenkel02a}.

\subsection{Dipolar spin ice}\label{DSI-physics}

\begin{figure}[ht]
\centering\includegraphics[width=10cm]{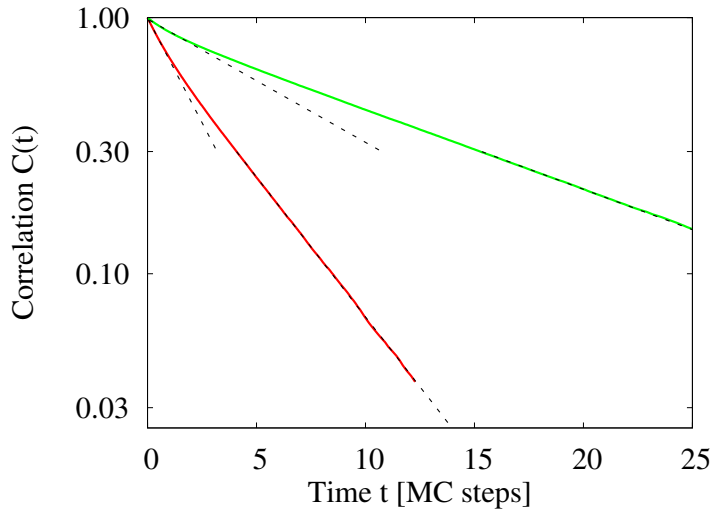}
\caption[$\;\;$ Time correlation function $C(t)$ for DSI] {\textbf{Time correlation function $C(t)$ for DSI} for $T=1.5$ (green) and 3 K (red) on a semi-log scale. The $x$-axis is the Monte Carlo time $t$. The dashed lines are exponential fits for short and long relaxation times.}
\label{fig:CtDyL10}
\end{figure}

Simulation of the DSI model is relatively straightforward. We implement both nearest neighbour and dipolar interactions, using the numerical values of $J$ and $D$ for Dy$_{2}$Ti$_{2}$O$_{7}$ given in table I.3. We extract the characteristic relaxation time $\tau$, from the auto-correlation function
\begin{eqnarray}
C(t)\;=\;\frac{1}{N}\sum_{i}\mathbf{S}_{i}(0)\cdot\mathbf{S}_{i}(t),
\label{eq:mono:autocorr1}
\end{eqnarray}
where $N$ is the total number of spins ($N=16\times8^{3}=8192$ here) and $\mathbf{S}_{i}(t)$ is the value of the unit vector representing the spin at the Monte Carlo time $t$. One Monte Carlo time step, for a system of size $N$ is defined as a single sweep through $N$ moments chosen at random. For the initial conditions we take a random configuration, which we let evolve at high temperature $T=10$ K until equilibrium, defined here as the time beyond which $C(t)$ decays below $0.01$. This defines $t=0$ and $C(t)$ is then computed and stored for a given temperature. The Monte Carlo time is re-set to zero when $C(t)$ decays beyond 0.001 and the temperature $T$ is lowered by $\delta T$. The process is repeated until we reach $T=0.7$ K when numerical equilibration becomes difficult. The resulting auto-correlation is averaged over 200 samples in order to give a smoothly decaying function down to $C(t)\lesssim 0.03$. The outcome is plotted in figure~\ref{fig:CtDyL10} for $T=1.5$ and 3 K.

$C(t)$ relaxes almost exponentially, but with a small difference between short and long time scales. Different time scales, $\tau_{i=1,2}$, can be extracted by fitting the correlation function $C(t)$ to an exponential function, either over the interval $C=1 - 0.8$ ( $\tau_{1}$) or over the interval  $C=0.3 - 0.03$ ($\tau_{2}$)(see dashed lines in figure~\ref{fig:CtDyL10}). The characteristic relaxation time $\tau$,
is taken to be the average of $\tau_{1}$ and $\tau_{2}$.
The Monte Carlo time is again scaled to that measured by Snyder \textit{\& al.}  by equating the scales at 4 K, $\tau(T=4\textrm{ K})= 2.99$ $ms$, for which the density of double defects is negligible ($\lesssim 1$\%), giving a quantitative equivalence between our Metropolis dynamics and the relaxation dynamics of the experiment, with  one Monte Carlo step $= 2.5 \pm 0.5$ ms.

\begin{figure} [ht]
\centering\includegraphics[scale=0.8]{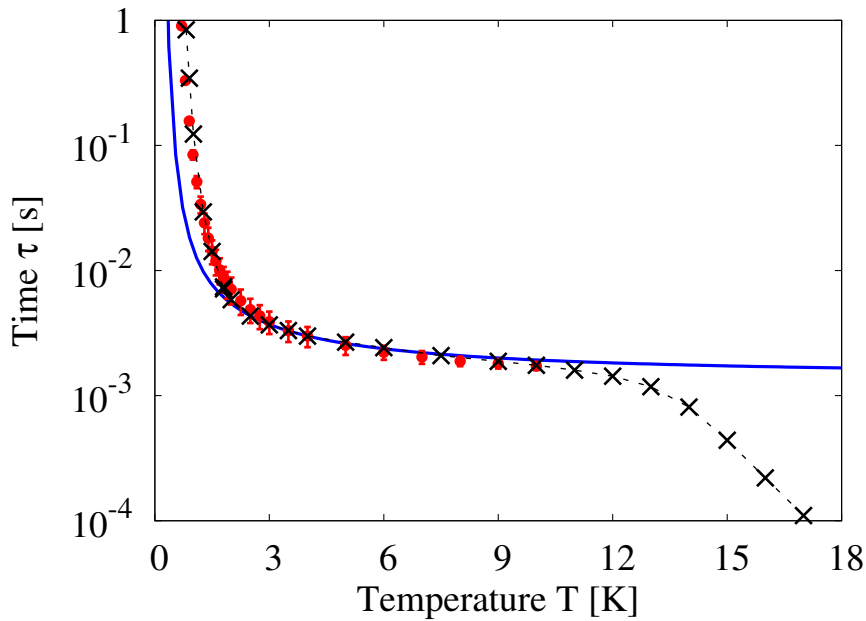}
\caption[$\;\;$ Relaxation time scale from DSI]{\textbf{Relaxation time scale from DSI} 
The experimental data ($\mathbf{\times}$) are from Snyder \textit{\& al.} \protect \cite{Snyder04b}
(\textcolor{red}{$\bullet$}). The system size is $N=16000$ spins. The details of the simulations are given in the text. 
The bars do not represent the error bars on the value of $\tau$; rather they show the extreme values $\tau_{1}$ and $\tau_{2}$ at each temperature.
In order to see the improvement due to dipolar interactions, the modified Arrhenius formula~\ref{eq:mono:Arr} is also plotted in blue.}
\label{fig:DSI-tau}
\end{figure}

Preliminary results (not shown) were for the NNSI model using the $J_{\rm eff}$ from Table 3. The data agree quantitatively with the analytic calculation,   $\tau_{0}\,\exp(\Delta E(T)/T)$, (see figure (\ref{fig:mono:Arr})) over the temperature range $0-10$ K showing that our rather simple approach accurately describes the stochastic dynamics of the NNSI model. Any deviation from this behaviour is therefore explicitly due to the long range interactions of the DSI. Our results for $\tau$ vs $T$ for the DSI are compared with the data of Snyder \textit{\& al.} \cite{Snyder04b} in figure (\ref{fig:DSI-tau}). We find remarkably good agreement between the numerical results  and experiment over the temperature range between 1 K and 10 K. In particular there is a substantial improvement over the Arrhenius law of figure (\ref{fig:mono:Arr}) below 3 K as we move into the low temperature regime with rapidly increasing time scale. Although when one looks more closely there are still some systematic differences between the data from the real and the numerical experiment, we believe that the general agreement shown in figure (\ref{fig:DSI-tau}) represents a considerable success for both the DSI and stochastic quantum tunneling hypothesis, in which the tunneling rate is temperature independent. Further, the tunneling time scale is approximately equal to the Monte carlo time scale fixed at 4 K, giving $\tau_0\sim 2.5$ miliseconds.

\begin{figure} [ht]
\centering\includegraphics[scale=0.8]{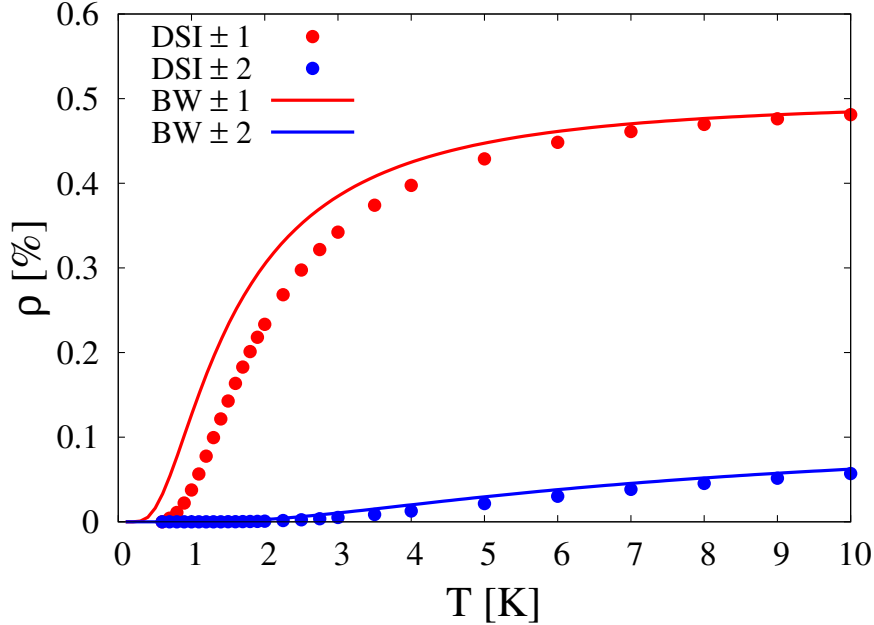}
\caption[$\;\;$ Density of monopoles]{\textbf{Density of monopoles} for the dipolar spin ice model ($\bullet$) and their expected Boltzmann weight for the nearest neighbour spin ice model (solid line) with the numerical values of Dy$_{2}$Ti$_{2}$O$_{7}$. In red (resp. blue) are plotted the concentration of \3 and \4 defects (resp. 4 \textit{in} and 4 \textit{out}), carrying a magnetic charge $\pm 1$ (resp. $\pm 2$). The system size is $N=16000$ spins.}
\label{fig:mono:dens}
\end{figure}

We start the investigation of exactly how the Coulomb interactions modify the spin ice dynamics by calculating the evolution of topological defects as a function of temperature in both the NNSI and DSI models. The evolution is shown in figure~\ref{fig:mono:dens}. In both cases the density of \3 and \4 defects shoots rapidly to zero below $1-2$ K, which corresponds to the low temperature regime,
while the double defect concentration goes to zero below the $5-6$ K range. This confirms the picture that the plateau region is due to the proliferation of a high density gas, while the low temperature region with increasing time scale is driven by an exponential fall in the monopole concentration. It is therefore clear from this picture that monopole physics should explain the agreement between real and numerical experiments shown in figure (\ref{fig:DSI-tau}). A second point to notice is that the monopole density falls to zero at a noticeable higher temperature for the DSI than for the NNSI model which must surely be related to  the brutal slowing down of the dynamics compared with our analytical Arrhenius law calculation. To investigate this slowing down further we move to the dumbbell model and a description of the problem uniquely in terms of monopoles. 

\subsection{Monopoles and strings}
\label{mono:def}

Within the monopole picture spin flips out of and into the constrained manifold correspond to creation/annihilation of pairs, while all other flips correspond to monopole hopping between nearest neighbour sites of the diamond lattice. Hence in a Monte Carlo step one considers two nearest neighbour sites at time $t$. The possible outcomes at time $t+1$ are:
\begin{itemize}
\item If there are no quasi-particles on these two sites, then we consider the creation of a pair of opposite charges.
\item If there is only one monopole, it can either move to the other site or stay where it is;
\item If there are two opposite charges, they can annihilate;
\item If there are two charges of the same sign, no movement is possible;  as we have disallowed monopole defects carrying double charges.
\end{itemize}

\subsubsection{Chemical potential $\mu$}

Moving into the monopole picture means we change the Gibbs ensemble from which we select microstates. In the magnetic picture the independent variables are $T$, magnetic field $\vec B$ (chosen to be zero throughout this paper) and the number of spins $N$, which is also fixed to the volume of the sample. In the monopole picture, the number of quasi particles varies, so the independent variables are $T$, $\vec B =0$, the volume $V$ and $\mu$ the chemical potential for monopole pair creation. The relevant thermodynamic potential is therefore the Grand Potential $\Phi = U -ST -N\mu$, for which we have to define $\mu$. We are, of course free to choose any value of $\mu$ and arbitrary values will take us away from spin ice physics, but as the present work is motivated by the experimental results of  Snyder {\it et. al.} we need to choose the value of $\mu$ that best corresponds to Dy$_2$Ti$_2$O$_7$. 
In a first series of simulations we have estimated $\mu=\mu_1$ numerically by calculating the difference between the Coulomb energy gained by creating a single pair of neighbouring magnetic monopoles of opposite sign $\Delta U_{\rm mono}<0$ in a vacuum and that required to produce a unique pair of topological defects out of the \2 manifold in the DSI, $\Delta U_{\rm def}>0$, giving a configurationally averaged estimate $\mu_{1}\equiv\Delta U_{\rm mono}-\Delta U_{\rm def} = -8.92$ K. Note that while $\Delta U_{\rm mono}$ is simply $V_{\rm min}$ from equation~(\ref{eq:mono:V2}), $\Delta U_{\rm def}$ has to be computed by simulation, as projective equivalence is not exact for the DSI. In a second series of simulations, $\mu_{2}(T)$ was taken as the value required to reproduce the same defect concentration as in a simulation of dipolar spin ice at temperature $T$, as shown in figure~\ref{fig:mono:dens}. We found that $\mu_2=\mu_{1}$ within numerical error at $T=0.7K$ and varied by  $\approx$3\% only, over the temperature range $1-4$ K, showing that our procedures are consistent and underlining the fact that the chemical potential used is \textit{not} a free parameter. Rather, it is taken from detailed comparison with the DSI that best describes Dy$_2$Ti$_2$O$_7$, with parameters from Table 3.  Note that in references \cite{Jaubert09a,Jaubert09c} we have used an unconventional definition of $\mu$, with a sign change compared to the above value. Here, we make a definition compatible with conventional statistical mechanics such that, in a Monte Carlo move we compute the change in $\delta{\tilde{U}} = \delta U_{mono} - \mu \delta N$. As the value given, $\mu\approx -8.9$ K, is for a pair of particles, $\delta N = \pm 1$ or zero.

\begin{figure} [ht]
\centering\includegraphics[scale=1.1]{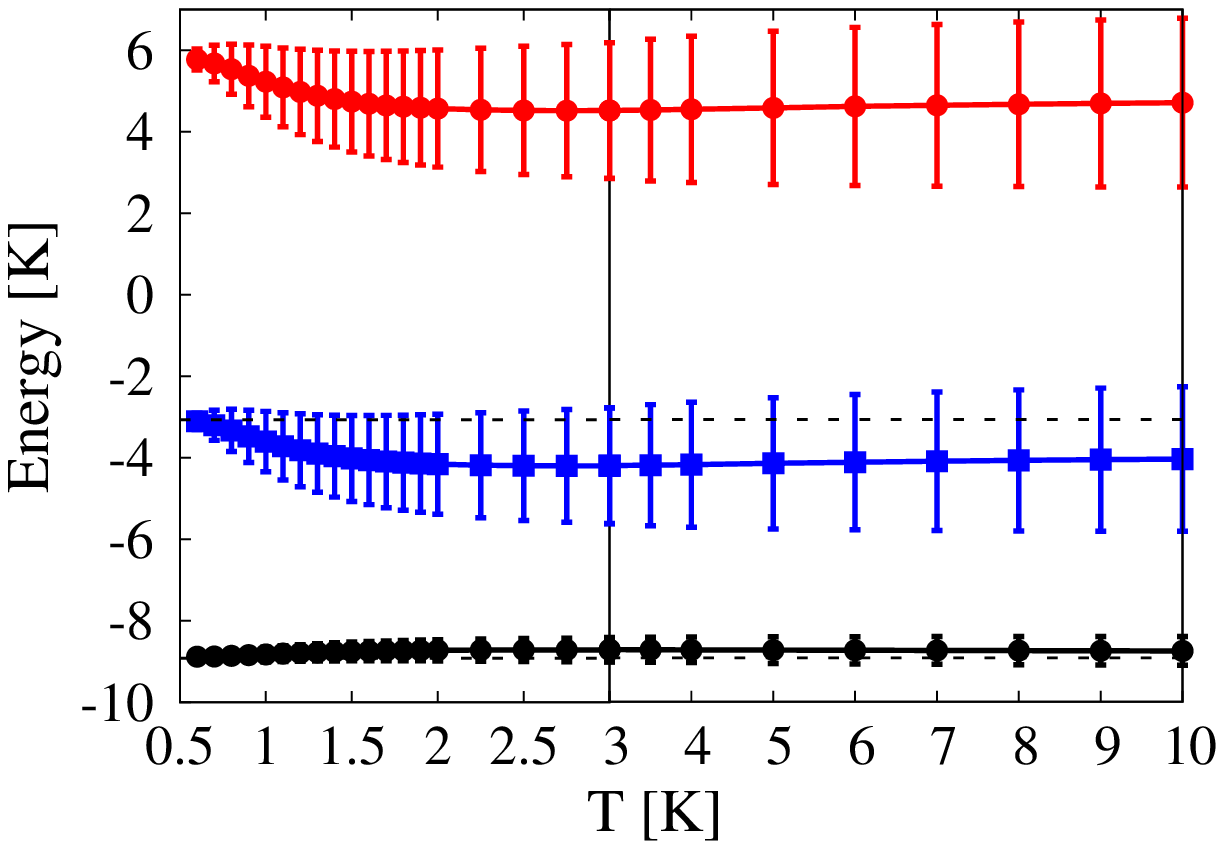}
\caption[$\;\;$ Creation of a pair of quasi-particles]{\textbf{Creation of a pair of quasi-particles:} Energy required in the dipolar spin ice model ($\Delta U_{def}$ , red $\textcolor{red}{\bullet}$) or gained for a Coulomb gas ($\Delta U_{mono}$, blue $\textcolor{blue}{\blacksquare}$) and the resulting chemical potential $\mu(T)$ ($\blacktriangle$) as a function of temperature. The vertical bars are \textit{not} error bars but the standard deviation of these quantities. $\mu(T)$ tends to the limit $\mu_{1}=-8.92$ K (lower dashed line) at very low temperature. The other dashed line is the analytical prediction of the energy gained by creating a pair of monopoles in a vacuum $V_{min}=-3.07$ K (see equation~(\ref{eq:mono:V1})). The temperature scale below 3 K is enlarged for a better display.}
\label{fig:mono:mu}
\end{figure}

It is interesting to dwell further on $\mu$ and its composite parts at finite temperature, as their evolution with $T$ turns out to be at the origin of the rapid increase in the time scale at low temperature. We can make a third estimate for the chemical potential; $\mu(T)=\Delta U_{\rm mono}(T) - \Delta U_{\rm def}(T)$ using the same procedure as for $\mu_1$, but at finite temperature, where a pair of monopoles is created in a microstate containing a finite concentration of monopoles corresponding to finite temperature $T$. To do this, we simulated the DSI model, with $N=8192$ spins  \textit{in parallel} with a 'slave' Coulomb gas, \ie for each single spin flip the corresponding monopole configuration on the diamond lattice is updated accordingly, making it possible to compare the energy of the spin configuration with its mirror in the dumbbell model at each Monte Carlo step. As we are interested in the creation of monopoles, we have artificially suppressed double defects in the DSI for all temperatures.
Hence after equilibration at a given temperature $T$ and for each accepted creation of a pair of defects, we compute the energy difference between the Monte Carlo move in the DSI model $\Delta U_{\rm def}$ and the evolution in the slave Coulomb gas $\Delta U_{\rm mono}$, which gives a value of  $\mu$ for this specific move. By averaging these values over the Monte Carlo time and for different initial configurations, one finds the temperature evolution of these energies as well as their standard deviation, plotted in figure~\ref{fig:mono:mu}. 

Much of the physics of spin ice at low temperature is included in this figure:  the standard deviation of both $\Delta U_{\rm def}(T)$ (red) and $\Delta U_{\rm mono}(T)$ (blue) are large above $1$ K,  up to $\pm 40 \%$ of the mean values. However, these quantities are  \textit{highly correlated} and their difference, giving rise to $\mu(T)$ (black) is very precisely defined with a standard deviation of $\pm 4 \%$ at most. Furthermore, the extracted $\mu(T)$  follows closely the values for $\mu_2$, which was estimated more phenomenologically. This agreement, despite the large fluctuations,
illustrates the success of the monopole picture and of the dumbbell model in describing the physics of spin ice: the energy of separation of two monopoles is given by the magnetic Coulomb potential, including many body effects for a finite particle concentration and the energy of creation (and annihilation, not shown here) is consistent with a Coulomb gas in the grand canonical ensemble. On average, large energy cost $\Delta U_{\rm def}$ (top of the red bars) correspond to small energy gain $\Delta U_{\rm mono}$ (top of the blue bars) and \textit{vice-versa}.

The data in figure (\ref{fig:mono:mu}) show two regimes of behaviour: from 10 K down to 2 K, corresponding to the plateau region for relaxation times and below 2 K where the relaxation time rapidly increases. In the higher temperature regime the mean values, $\langle\Delta U_{\rm def}\rangle$  and $\langle\Delta U_{\rm mono}\rangle$ are roughly constant and   $\left<\Delta U_{\rm def}\right>= \langle\Delta U_{\rm mono}\rangle-\langle \mu(T)\rangle $ is close to the energy cost of a single spin flip in the NNSI model $E=4\,J_{\textrm{eff}}=4.44$ K. Thus, apart from the fact that there is a quasi-continuous range of energy values to create monopoles instead of a unique one and that the fluctuations in the energy scales are surprisingly large, the results are consistent with the NNSI model, which explains why the NNSI provides a reasonable description of spin ice down to 2 K. This equivalence between models with long and short range interactions is a direct consequence of Debye screening \cite{Morris09a,Henley10a}. That is, throughout this regime, $\langle\Delta U_{\rm mono}\rangle$ (blue curve) is in excess of the energy of deconfinement of a pair of monopoles in a vacuum, $V_{min}=-3.07$ K (lower dashed line, see equation~(\ref{eq:mono:V1})) and from the fluctuations one can see that pair creations providing less energy are rare. 
Hence, a newly created pair is more stable than in a vacuum with the additional energy coming from the build up of correlations within the Coulomb gas, favouring a neighbouring charge cloud of opposite charge around each particle. As the temperature is reduced, screening becomes more marked so that low energy events become increasingly rare, leading a maximum in the mean Coulomb energy gain on pair creation, at around 2.5 K. However, below this temperature the monopole concentration drops rapidly (see figure \ref{fig:mono:dens}) so that the Debye screening is starved and there is a marked evolution of the energies in figure (\ref{fig:mono:mu}).
This appears to occur as the smallest values of $\Delta U_{\rm mono}$ disappear brutally and $\langle\Delta U_{\rm mono}\rangle$ increases rapidly towards $V_{min}$. This evolution is accompanied by a 
surprising increase in $\langle\Delta U_{\rm def}\rangle$ (red line), ensuring that $\mu(T)$ remains temperature independent. The \textit{increase} in the mean energy required to create a pair of topological defects as temperature \textit{decreases} is clearly due to the absence of many body screening effects from the mobile quasi-particles. 
 The brutal non-Arrhenius spin freezing as one leaves the plateau region at low temperature is therefore an \textbf{\textit{avalanche effect}}: fewer monopoles hinders the Debye screening, which reduces the number of creations of quasi-particles, which leads to less monopoles, ... If this slowing down is sharp \textit{but} continuous, it is because the system remains thermally activated. That is, as long as the material remains ``hot'' enough to allow the creation of a pair of monopoles out of the vacuum,\ie a single spin flip of energy $\Delta U_{\rm def}=\mu_{1}+V_{min}\approx 5.8$ K, the system will remain dynamic through the creation and diffusion of monopoles. Our simulations hit this limit for $T\approx 0.6$ K and one might expect equilibration in experiments to become dramatically difficult below this temperature. This is exactly what Snyder \textit{\& al.} observed at 0.65 K for Dy$_{2}$Ti$_{2}$O$_{7}$ in FC-ZFC measurements \cite{Snyder04b}, although  magnetocaloric measurements \cite{Orendac07a} and recent muon spin resonance experiments\cite{Bramwell09a} appear to measure the dynamical properties of spin ice materials at considerably lower temperature where the intrinsic monopole concentration, as predicted by the DSI or the dumbbell model would be considerably lower.

\subsubsection{Dirac strings}

 The monopoles hop between nearest neighbour sites via the Metropolis Monte Carlo algorithm, giving diffusive dynamics, but with the local constraint discussed in the introduction: in the spin model a \3 topological defect can move at low energy cost by flipping one of the three \textit{in} spins, the direction of the \textit{out} spin being barred by an energy barrier of $8\,J_{\textrm{eff}}$ that would result in a double defect and which is set to infinity in the Coulomb gas. An isolated monopole can therefore hop to 3 out of 4 of its nearest neighbour sites only, dictated by an oriented network of constrained trajectories corresponding to the ensemble of classical \textit{Dirac strings} \cite{Castelnovo08a} of connected dipoles \cite{Jackson99a}. The positively charged monopoles move in one sense along the network while the negative charges move in the opposite direction (see figure~\ref{fig:mono:mono}.b). The network is dynamically re-arranged through the evolution of the monopole configuration.  In fact the characteristic time scale that we compare with experiment comes from the evolution of the network of Dirac strings rather than from the monopoles themselves.  That is, the magnetic relaxation experiment of Snyder {\it et. al.} \cite{Snyder04b} ultimately measures spin dynamics, due monopole diffusion, rather than giving a direct measure of the  monopoles themselves.
We define the string network by an integer $\sigma=\pm 1$ giving the orientation of the Dirac string along each bond of the diamond lattice. The Dirac string network is an image of the underlying spin structure for which we can define the
auto-correlation function
\begin{eqnarray}
C(t)\;=\;\frac{1}{N}\sum_{i}\sigma_{i}(0)\sigma_{i}(t).
\label{eq:mono:autocorr2}
\end{eqnarray}
We start the simulation with the Dirac string network in an ordered configuration and let it evolve
at $T=4$ K until equilibrium is reached. 
A high density of monopoles is generated, which move following the criteria defined above and time step is defined as a sweep through $N$ links of the monopole vacuum.
We  compute $C(t)$ averaged over $200$ simulations following the same procedure as for the calculation of spin-spin autocorrelation function in the DSI. The Monte Carlo time was again scaled to the experimental time at $T=4$ K and no other fitting parameters were used.

\subsection{Comparison with experiment}

\begin{figure}[ht]
\centering\includegraphics[scale=0.8]{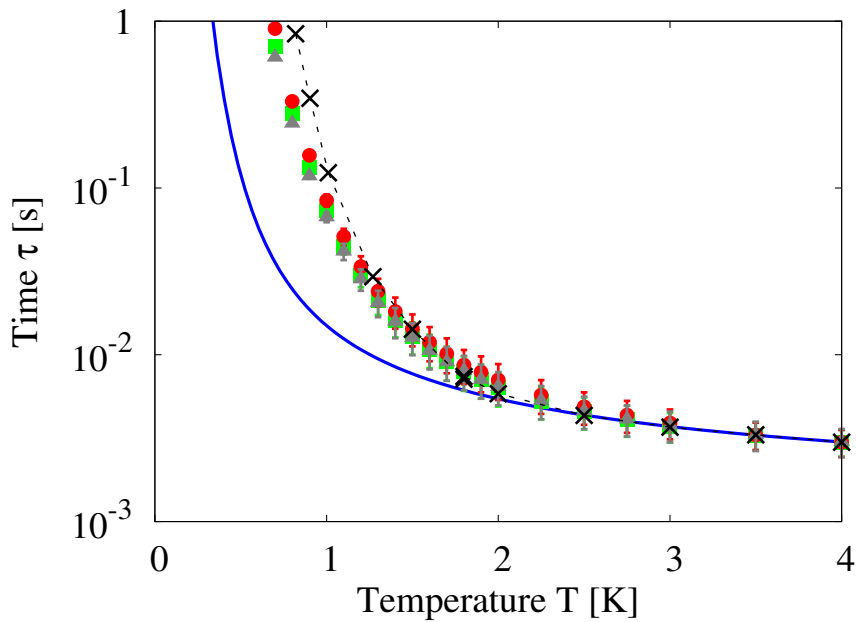}
\caption[$\;\;$ Spin freezing in Dy$_{2}$Ti$_{2}$O$_{7}$]{\textbf{Spin freezing in Dy$_{\mathbf{2}}$Ti$_{\mathbf{2}}$O$_{\mathbf{7}}$:} The experimental data ($\mathbf{\times}$) are from Snyder \textit{\& al.} \protect \cite{Snyder04b}. The Arrhenius law (blue line) represents the free diffusion of topological defects in the nearest neighbour model. The relaxation time scale of the Dirac string network for the Coulomb gas has been obtained for fixed chemical potential (\textcolor{gray}{$\blacktriangle$}) and with $\mu$ varying slowly to match the defect concentration in dipolar spin ice (\textcolor{green}{$\blacksquare$}), whereas the dipolar spin ice relaxation time is given by \textcolor{red}{$\bullet$}. As in figure~\ref{fig:DSI-tau} and as defined in the text, the bars do not represent the error bars on the value of $\tau$; rather they show the extreme values $\tau_{1}$ and $\tau_{2}$ at each temperature. The system size is $8\times 12^{3}=13824$ sites on the diamond lattice for the Coulomb gas and 8192 spins on the pyrochlore lattice for DSI simulations.}
\label{fig:mono:tau}
\end{figure}

In accordance with our analysis of figure~\ref{fig:mono:mu}, the characteristic time scale for relaxation of the Dirac string network does indeed reproduce the sharp spin freezing observed experimentally \cite{Jaubert09a}. Results are shown in figure (\ref{fig:mono:tau}) in the interval below 4 K. Data from the Coulomb gas simulation is in very close agreement with the Monte Carlo data for the DSI, confirming that the low temperature dynamics of the DSI is driven by the dynamics of Coulombic quasi particles, so that a Coulomb gas in the grand canonical ensemble really is the correct description of the system at low temperature. Also shown in the figure is the modified Arrhenius formula~\ref{eq:mono:Arr} describing freezing in the NNSI.  On this temperature scale the difference between data for the NNSI and that for the DSI or the Coulomb gas is considerable, confirming that the sharp, non-Arrhenius freezing at low temperature is due to the presence of long range interactions between  quasi-particles. The most significant effect here is the gradual suppression of Debye screening below $\sim 2$ K, provoking the avalanche effect which accelerates the disappearance of the monopoles below this temperature. In addition we also expect the diffusive dynamics to be slowed by the formation of neutral pairs, bound by the Coulomb interaction. However, as the Coulomb interaction is not confining in three dimensions the monopole mobility should remain finite. Below $1.5$ K the estimate of the time scale for the Coulomb gas lies slightly below that for the DSI. This small difference provides a measure of the approximation involved in passing from the spin system to the charged particles. 

We show simulation results using both our definitions of the chemical potential. The data sets are very close to each other, although a small systematic difference appears around 1 K, with data for the fixed chemical potential, $\mu_1$ giving a slightly lower estimate for $\tau$. This small deviation seems consistent with the small evolution in $\mu(T)$ vs $T$ observed in figure (\ref{fig:mono:mu}). 

Finally, when looked at with this level of precision, all numerical data falls below the experimental data at low temperature. This small but systematic difference may indicate the limits of applicability of both the DSI and our model of stochastic dynamics although more experiments need to be done to quantify this remark. More recent analysis of the static properties of Dysprosium Titanate suggest that the DSI can be improved by including further neighbour exchange\cite{Yavorskii08a}, which will change fine details of the dynamics. The approximation of having a single, temperature independent microscopic time scale, $\tau_0$, will also have its limitations which will also be manifested in the fine detail of the difference between simulation and experiment. 

\section{Conclusion}

The recent discovery of monopole quasi-particles in spin ice models and materials provides a rare occasion to think outside the magnetic box, while studying a set of magnetic materials. The theoretical\cite{Castelnovo08a, Morris09a}, numerical\cite{Jaubert09a,Jaubert09c,Castelnovo10b} and experimental\cite{Fennell09a,Morris09a,Bramwell09a,Kadowaki09a} work so far presented in this field make this point very clear. Here, the collective quasi-particle excitations are localized in real space, rather than reciprocal space, despite the absence of quenched disorder. The quasi-particle vacuum has hidden internal degrees of freedom with $U(1)$ gauge symmetry and thanks to the unique role played by the dipole interactions on the pyrochlore lattice structure, the excitations have both the topological and energetic properties of magnetic charges. The correct description at low temperature is therefore that of a Coulomb (lattice) gas in the grand canonical ensemble, whose properties are almost identical to those of its electronic equivalent, an electrolyte of positive and negatively charged ions whose dynamics can be considered as purely stochastic.

There are some exceptions to this analogy, notably spin ice materials cannot support a $dc$ monopole current in the steady state\cite{Jaubert09a}. This is because of the dynamical constraint that positive and negative charges move in opposite directions over the underlying Dirac string network, changing the direction of the network as they pass. The system only posses a finite number of monopole trajectories in each direction before complete ordering of the network occurs. Or, in magnetic language, the sample can provide a finite number of spin flips in a given direction before becoming magnetically ordered. This fact seems difficult to reconcile with recent experiments\cite{Bramwell09a}, where the magnetic charge has been measured through application of Onsager's Wien effect \cite{Onsager34a} to Dysprosium Titanate. In an electrolyte, in the low charge concentration limit,  the electric current density, $\sigma(E)$ increases linearly with electric field $E$, with the result that the ratio $\sigma(E)/\sigma(0)$ is a universal number containing $E$, $T$, charge $q$ and fundamental constants only. By treating muon relaxation rate in an equivalent way to the electric current, Bramwell {\it et. al.}  have been able to measure the magnetic charge $Q\approx 4.6 \mu_B$ A$^{-1}$ in good agreement with the theoretical prediction of Castelnovo {\it et. al.} \cite{Castelnovo08a}. How can this be, given that no permanent current can exist ? One possible explanation is that, in weak fields and on short time scales, the constraints do not qualitatively affect the dynamics. One can define a drift length scale for  field $X$, $\ell_X = k_BT/QX$, where $X$ is the magnitude of a magnetic or electric field. Above this scale one should observe the deterministic drift of the ions in the electrolyte and the effects of the constraints in the magnetic problem. In the field and temperature range of the Wien effect experiment ($T=100-200$ mK, $B \approx 2$ mT), $\ell_B\sim 70$ \AA. Given this large scale it seems reasonable that the magnetic and electric systems should show the same behaviour within the experimental set up. Much work is required here to understand these points in detail.

Spin ice dynamics at low temperature should be compared with the dynamics of a spin glass. In the latter, the glassy behaviour is the result of a hierarchy of time scales, as manifested, for example in the phenomenological random energy model \cite{Derrida81a}. Here the evolution steps occur with an exponentially varying set of delay times. For spin ice, at least in zero field things are very different. 
We have modeled the dynamics of the NNSI using a single time scale associated with the quantum tunneling of an Ising spin $\tau_0$. This scale dictates two processes; a hopping move of an existing deconfined monopole by one lattice spacing, or the creation (annihilation) of a nearest neighbour pair of particles, whose characteristic time is dressed by the Boltzmann weight, $\tau=\tau_o\exp(-\beta\Delta E)$. At low temperature this becomes the rate determining step, so that once created, the diffusing particles are actually extremely efficient at decorrelating the system. For example, from figure (\ref{fig:mono:Arr}), the characteristic Arrhenius time scale at 500 mK is $\tau\sim 1$ sec, while the microscopic time scale, $\tau_0$ is estimated to be about $2.5 $ miliseconds. That is, the system decorrelates in $400$ microscopic time steps thanks to the free diffusion of the deconfined particles. In the presence of Coulomb interactions, the decorrelation times are longer as the microscopic processes are dressed by a continuous range of energy scales coming from the Coulomb interaction, but the basic picture is provided by the simpler nearest neighbour model where free particles move on an essentially flat energy surface, making them efficient in decorrelating the system. 

In this paper we have interpreted a magnetic relaxation experiment in terms of monopole dynamics. A more fundamental approach is to design experiments that explicitly and directly test for monopole dynamics. Up to press the only experiments which attempt this are those using the Wien effect\cite{Bramwell09a,Giblin10a}, in which monopole mobility can be probed by exploiting the chemical equilibrium existing between bound and free particles in a field. An even more direct approach would be an equivalent of the  ``Stanford'' superconducting coil experiment \cite{Cabrera82a,Castelnovo08a}, whose goal was to measure the passage of free Dirac monopoles through a superconducting coil. Such an experiment could in principle detect the passage of a single magnetic quasi-particle but, given that the charges have no mass and therefore have diffusive, rather than Newtonian dynamics, at least in the absence of an external field and that the monopole charge in Dysprosium Titanate is approximately 8000 times smaller than a Dirac Monopole\cite{Castelnovo08a}, this experiment presents serious challenges. A starting point could be a many body equivalent. In fact, such an experiment  has been set up in another context by H\'errison and Ocio\cite{Herisson02a} to look at the violation of the fluctuation dissipation relation in spin glass systems. In this experiment the magnetic fluctuations are directly measured and compared with magnetic response. The analysis presented in this paper suggests that the resulting magnetic noise in spin ice would be due to the diffusive motion of monopoles.  Unfortunately, for the moment this set up is limited to the 4 K temperature range which remains in the high monopole density regime. The goal for a monopole experiment would be first to descend below one Kelvin and then to reduce the scale of the sample sufficiently so that the resulting noise was due to a mesoscopic, or even microscopic number of quasi-particles.

In conclusion, the discovery of magnetic monopoles in spin ice has exposed a rich vein of physics, in which a Coulomb gas, equivalent to a weak electrolyte appears at low temperature. This emergent 'magnetolyte'\cite{Bramwell09a} shows many of the static and dynamic properties of the electric system within a very idealized environment. Much experimental, theoretical and numerical work remains to be done in order to fully exploit it. 

\section*{Acknowledgements}

It is a pleasure to thank S. T. Bramwell and R. Moessner for stimulating discussions and for related collaborations. We have, in addition had useful conversations with L. Balents, L. Berthier, B. Canals,  C. Castelnovo, T. Fennell, S. Giblin, M. Gingras,  V. Kaiser, K. Matsuhira and P. Schiffer. PCWH thanks the Max Planck Institute for Complex Systems in Dresden for financial support through the study program, "Unconventional Magnetism in High Fields" and apologises to the organizers for not exceeding 10 mT.

\bibliographystyle{plain}

\end{document}